\documentclass[usenatbib,fleqn,longauth]{aa}  

\usepackage{graphicx}
\usepackage{txfonts}

\usepackage{euclid}
\usepackage{newtxtext,bm}
\usepackage{amsmath}
\usepackage[T1]{fontenc}
\usepackage{ae,aecompl}
\usepackage[colorinlistoftodos,backgroundcolor=pink]{todonotes}
\usepackage{tabularx}
\usepackage{tikz}
\usepackage[normalem]{ulem}
\usepackage{color, colortbl}
\definecolor{RED}{rgb}{0.9,.5,.5}
\usetikzlibrary{bayesnet}
\defcitealias{Massey13}{M13}
\defcitealias{Cropper13}{C13}
\usetikzlibrary{shapes.geometric, arrows}
\usetikzlibrary{shapes,arrows}
\def\gs{\mathrel{\raise1.16pt\hbox{$>$}\kern-7.0pt %
\lower3.06pt\hbox{{$\scriptstyle \sim$}}}}         %
\def\ls{\mathrel{\raise1.16pt\hbox{$<$}\kern-7.0pt %
\lower3.06pt\hbox{{$\scriptstyle \sim$}}}}         %
\usepackage{graphicx}	
\usepackage{amsmath}	
\usepackage{amssymb}	

\renewcommand{\d}{\mathrm{d}}

\newcommand{\bee}{\begin{equation}}
\newcommand{\eee}{\end{equation}}
\newcommand{\nn}{\nonumber\\}
\newcommand{\sersic}{S\'{e}rsic}

\begin{document} 
 
\title{\Euclid preparation: VI. Verifying the Performance of Cosmic Shear Experiments}

\author{\Euclid Collaboration, P.~Paykari$^{1}$, T. D.~Kitching$^{1}$\thanks{E-mail: t.kitching@ucl.ac.uk}, H.~Hoekstra$^{2}$, R.~Azzollini$^{1}$, 
V.F.~Cardone$^{3}$, M.~Cropper$^{1}$, C.A.J.~Duncan$^{4}$, A.~Kannawadi$^{2}$, L.~Miller$^{4}$, H.~Aussel$^{5}$
% ...
}

\institute{$^{1}$ Mullard Space Science Laboratory, University College London, Holmbury St Mary, Dorking, Surrey RH5 6NT, UK\\
$^{2}$ Leiden Observatory, Leiden University, Niels Bohrweg 2, 2333 CA Leiden, The Netherlands\\
$^{3}$ INAF-Osservatorio Astronomico di Roma, Via Frascati 33, I-00078 Monteporzio Catone, Italy\\
$^{4}$ Department of Physics, Oxford University, Keble Road, Oxford OX1 3RH, UK\\
$^{5}$ AIM, CEA, CNRS, Universit\'{e} Paris-Saclay, Universit\'{e} Paris Diderot, Sorbonne Paris Cit\'{e}, F-91191 Gif-sur-Yvette, France\\
$^{6}$ Department of Physics, University of Malta, Msida, MSD 2080, Malta\\
$^{7}$ Institute of Space Sciences and Astronomy (ISSA), University of Malta, Msida, MSD 2080, Malta\\
$^{8}$ INAF-Osservatorio di Astrofisica e Scienza dello Spazio di Bologna, Via Piero Gobetti 93/3, I-40129 Bologna, Italy\\
$^{9}$ Dipartimento di Fisica e Astronomia, Universit\'a di Bologna, Via Gobetti 93/2, I-40129 Bologna, Italy\\
$^{10}$ INFN-Sezione di Bologna, Viale Berti Pichat 6/2, I-40127 Bologna, Italy\\
$^{11}$ INAF-Osservatorio Astronomico di Trieste, Via G. B. Tiepolo 11, I-34131 Trieste, Italy\\
$^{12}$ INAF-Osservatorio Astrofisico di Torino, Via Osservatorio 20, I-10025 Pino Torinese (TO), Italy\\
$^{13}$ Dipartimento di Fisica e Astronomia “G.Galilei", Universit\'a di Padova, Via Marzolo 8, I-35131 Padova, Italy\\
$^{14}$ Department of Astronomy, University of Geneva, ch. d'\'Ecogia 16, CH-1290 Versoix, Switzerland\\
$^{15}$ INFN-Sezione di Roma Tre, Via della Vasca Navale 84, I-00146, Roma, Italy\\
$^{16}$ Department of Mathematics and Physics, Roma Tre University, Via della Vasca Navale 84, I-00146 Rome, Italy\\
$^{17}$ Institut de Recherche en Astrophysique et Plan\'etologie (IRAP), Universit\'e de Toulouse, CNRS, UPS, CNES, 14 Av. Edouard Belin, F-31400 Toulouse, France\\
$^{18}$ INAF-Osservatorio Astronomico di Capodimonte, Via Moiariello 16, I-80131 Napoli, Italy\\
$^{19}$ Instituto de Astrof\'isica e Ci\^encias do Espa\c{c}o, Universidade do Porto, CAUP, Rua das Estrelas, PT4150-762 Porto, Portugal\\
$^{20}$ INFN-Bologna, Via Irnerio 46, I-40126 Bologna, Italy\\
$^{21}$ Dipartimento di Fisica e Scienze della Terra, Universit\'a degli Studi di Ferrara, Via Giuseppe Saragat 1, I-44122 Ferrara, Italy\\
$^{22}$ INAF, Istituto di Radioastronomia, Via Piero Gobetti 101, I-40129 Bologna, Italy\\
$^{23}$ INFN-Sezione di Torino, Via P. Giuria 1, I-10125 Torino, Italy\\
$^{24}$ Dipartimento di Fisica, Universit\'a degli Studi di Torino, Via P. Giuria 1, I-10125 Torino, Italy\\
$^{25}$ INAF-IASF Milano, Via Alfonso Corti 12, I-20133 Milano, Italy\\
$^{26}$ Institut de F\'isica d'\'Altes Energies IFAE, 08193 Bellaterra, Barcelona, Spain\\
$^{27}$ Institut de Ciencies de l'Espai (IEEC-CSIC), Campus UAB, Carrer de Can Magrans, s/n Cerdanyola del Vall\'es, 08193 Barcelona, Spain\\
$^{28}$ Institut d'\'Estudis Espacials de Catalunya (IEEC), 08034 Barcelona, Spain\\
$^{29}$ Department of Physics "E. Pancini", University Federico II, Via Cinthia 6, I-80126, Napoli, Italy\\
$^{30}$ INFN section of Naples, Via Cinthia 6, I-80126, Napoli, Italy\\
$^{31}$ INAF-Capodimonte Observatory, Salita Moiariello 16, I-80131, Napoli, Italy\\
$^{32}$ Aix-Marseille Univ, CNRS, CNES, LAM, Marseille, France\\
$^{33}$ Centre National d'Etudes Spatiales, Toulouse, France\\
$^{34}$ Instituto de Astrof\'{i}sica de Canarias. Calle V\'{i}a L\`{a}ctea s/n, 38204, San Crist\'{o}bal de la Laguna, Tenerife, Spain\\
$^{35}$ Institute for Astronomy, University of Edinburgh, Royal Observatory, Blackford Hill, Edinburgh EH9 3HJ, UK\\
$^{36}$ University of Nottingham, University Park, Nottingham NG7 2RD, UK\\
$^{37}$ ESAC/ESA, Camino Bajo del Castillo, s/n., Urb. Villafranca del Castillo, 28692 Villanueva de la Ca\~nada, Madrid, Spain\\
$^{38}$ Universit\'e de Lyon, F-69622, Lyon, France ; Universit\'e de Lyon 1, Villeurbanne; CNRS/IN2P3, Institut de Physique Nucl\'eaire de Lyon, France\\
$^{39}$ Departamento de F\'isica, Faculdade de Ci\^encias, Universidade de Lisboa, Edif\'icio C8, Campo Grande, PT1749-016 Lisboa, Portugal\\
$^{40}$ Instituto de Astrof\'isica e Ci\^encias do Espa\c{c}o, Faculdade de Ci\^encias, Universidade de Lisboa, Campo Grande, PT-1749-016 Lisboa, Portugal\\
$^{41}$ Department of Physics \& Astronomy, University of Sussex, Brighton BN1 9QH, UK\\
$^{42}$ Institut d'Astrophysique Spatiale, Universite Paris-Sud, Batiment 121, 91405 Orsay, France\\
$^{43}$ Institute of Space Sciences (ICE, CSIC), Campus UAB, Carrer de Can Magrans, s/n, 08193 Barcelona, Spain\\
$^{44}$ Aix-Marseille Univ, CNRS/IN2P3, CPPM, Marseille, France\\
$^{45}$ Max Planck Institute for Extraterrestrial Physics, Giessenbachstr. 1, D-85748 Garching, Germany\\
$^{46}$ Universit\"ats-Sternwarte M\"unchen, Fakult\"at f\"ur Physik, Ludwig-Maximilians-Universit\"at M\"unchen, Scheinerstrasse 1, 81679 M\"unchen, Germany\\
$^{47}$ von Hoerner \& Sulger GmbH, Schlo{\ss}Platz 8, D-68723 Schwetzingen, Germany\\
$^{48}$ Max-Planck-Institut f\"ur Astronomie, K\"onigstuhl 17, D-69117 Heidelberg, Germany\\
$^{49}$ Department of Physics and Helsinki Institute of Physics, Gustaf H\"allstr\"omin katu 2, 00014 University of Helsinki, Finland\\
$^{50}$ Institut d'Astrophysique de Paris, 98bis Boulevard Arago, F-75014, Paris, France\\
$^{51}$ Institut de Physique Nucl\'eaire de Lyon, 4, rue Enrico Fermi, 69622, Villeurbanne cedex, France\\
$^{52}$ Universit\'e de Gen\`eve, D\'epartement de Physique Th\'eorique and Centre for Astroparticle Physics, 24 quai Ernest-Ansermet, CH-1211 Gen\`eve 4, Switzerland\\
$^{53}$D\'{e}partement de Physique Th\'{e}orique and Center for Astroparticle Physics, Universit\'{e} de Gen\`{e}ve, 24 quai Ernest Ansermet, CH-1211 Geneva, Switzerland\\
$^{54}$ European Space Agency/ESTEC, Keplerlaan 1, 2201 AZ Noordwijk, The Netherlands\\
$^{55}$ Institute of Theoretical Astrophysics, University of Oslo, P.O. Box 1029 Blindern, N-0315 Oslo, Norway\\
$^{56}$ Argelander-Institut f\"ur Astronomie, Universit\"at Bonn, Auf dem H\"ugel 71, 53121 Bonn, Germany\\
$^{57}$Stuto Nazionale di Astrofisica (INAF) - Osservatorio di Astrofisica e Scienza dello Spazio (OAS), Via Gobetti 93/3, I-40127 Bologna, Italy\\
$^{58}$ Centre for Extragalactic Astronomy, Department of Physics, Durham University, South Road, Durham, DH1 3LE, UK\\
$^{59}$ INAF-Osservatorio Astronomico di Padova, Via dell'Osservatorio 5, I-35122 Padova, Italy\\
$^{60}$ University of Paris Denis Diderot, University of Paris Sorbonne Cit\'e (PSC), 75205 Paris Cedex 13, France\\
$^{61}$ IRFU, CEA, Universit\'e Paris-Saclay F-91191 Gif-sur-Yvette Cedex, France\\
$^{62}$ Sorbonne Universit\'e, Observatoire de Paris, Universit\'e PSL, CNRS, LERMA, F-75014, Paris, France\\
$^{63}$ INAF-IASF Bologna, Via Piero Gobetti 101, I-40129 Bologna, Italy\\
$^{64}$ Institute of Cosmology and Gravitation, University of Portsmouth, Portsmouth PO1 3FX, UK\\
$^{65}$ Kapteyn Astronomical Institute, University of Groningen, PO Box 800, 9700 AV Groningen, The Netherlands\\
$^{66}$ Space Science Data Center, Italian Space Agency, via del Politecnico snc, 00133 Roma, Italy\\
$^{67}$School of Physics and Astronomy, Queen Mary University of London, Mile End Road, London E1 4NS, UK\\
$^{68}$ INFN-Padova, Via Marzolo 8, I-35131 Padova, Italy\\
$^{69}$ Jet Propulsion Laboratory, California Institute of Technology, 4800 Oak Grove Drive, Pasadena, CA, 91109, USA\\
$^{70}$ Departamento de F\'isica, FCFM, Universidad de Chile, Blanco Encalada 2008, Santiago, Chile\\
$^{71}$ I.N.F.N.-Sezione di Roma Piazzale Aldo Moro, 2 - c/o Dipartimento di Fisica, Edificio G. Marconi, I-00185 Roma, Italy\\
$^{72}$ Centro de Investigaciones Energ\'eticas, Medioambientales y Tecnol\'ogicas (CIEMAT), Avenida Complutense 40, 28040 Madrid, Spain\\
$^{73}$ Instituto de Astrof\'isica e Ci\^encias do Espa\c{c}o, Faculdade de Ci\^encias, Universidade de Lisboa, Tapada da Ajuda, PT-1349-018 Lisboa, Portugal\\
$^{74}$ Universidad Polit\'ecnica de Cartagena, Departamento de Electr\'onica y Tecnolog\'ia de Computadoras, 30202, Cartagena, Spain\\
$^{75}$ Universit\'{e} C\^{o}te d'Azur, Observatoire de la C\^{o}te d'Azur, CNRS, Laboratoire Lagrange, Bd de l'Observatoire, CS 34229, 06304 Nice cedex 4, France\\
$^{76}$ CEICO, Institute of Physics of the Czech Academy of Sciences, Na Slovance 2, Praha 8 Czech Republic\\
$^{77}$ IRAP, Université de Toulouse, CNRS, CNES, UPS, Toulouse, France}
  \abstract
   {}
   {Our aim is to quantify the impact of systematic effects on the inference of cosmological parameters from cosmic shear.}
   {We present an `end-to-end' approach that introduces sources of bias in a modelled weak lensing survey on a galaxy-by-galaxy level. Residual biases are propagated through a pipeline from galaxy properties (one end) through to cosmic shear power spectra and cosmological parameter estimates (the other end), to quantify how imperfect knowledge of the pipeline changes the maximum likelihood values of dark energy parameters.}
   {We quantify the impact of an imperfect correction for charge transfer inefficiency (CTI) and modelling uncertainties of the point spread function (PSF) for \emph{Euclid}, and find that the biases introduced can be corrected to acceptable levels.}
   {}


\titlerunning{Euclid preparation: VI.}
\authorrunning{Euclid Collaboration et al.}

\maketitle
%
\section{Introduction}
Over the past century advances in observational techniques in cosmology have led to a number of important discoveries, of which the accelerating expansion of the Universe is perhaps the most surprising. Moreover, a wide range of detailed observations can be described with a model that requires a remarkably small number of parameters, which have been constrained with a precision that was unimaginable only thirty years ago.  This `concordance' model, however, relies on two dominant ingredients of the mass-energy content of the Universe: dark matter and dark energy, neither of which can be described satisfactorily by our current theories of 
particle physics and gravity. Although a cosmological constant/vacuum energy is an excellent fit to the current data, 
the measured value appears to be unnaturally small. Many alternative 
explanations have been explored, including modifications of the theory of General Relativity on large scales
\citep[see e.g.][for a review]{Amendola13}, but a more definitive solution may require observational constraints that 
are at least an order of magnitude more precise.  

The concordance model can be tested by studying the expansion history of the Universe and by determining the rate 
at which structures grow during this expansion. This is the main objective of the {\it Euclid} 
mission \citep{Laureijs11}, which will carry out a survey of 15\,000 deg$^2$ of the extragalactic sky. 
Although {\it Euclid} will enable a wide range of science topics, it is designed with two main probes 
in mind: (i) the measurement of the clustering of galaxies at $z>0.9$ using near-infrared, 
slitless spectroscopy; (ii) the direct measurement of the distribution of matter as a function 
of redshift using weak gravitational lensing, the effect whereby coherent shear distortions 
in the images of distant galaxies are caused by the differential deflection of light by intervening large-scale structures. The two-point statistics of the weak gravitational lensing caused by large-scale structure is known as `cosmic shear' \citep[see][for a recent review]{Kilbinger15}.
In this paper we explore the impact of instrumental effects and scanning strategy on the accuracy and precision with which dark energy parameters $w_0$ and $w_a$ \citep{Chevallier_Polarski_2001,Linder_2003} can be measured using the cosmic shear from {\it Euclid}. 

The challenge of measuring the cosmic shear signal is that the typical change in polarisation (third flattening or eccentricity) caused by gravitational lensing is approximately one percent, much smaller than the intrinsic (unlensed) ellipticities of galaxies. 
To overcome this source of statistical uncertainty, cosmic shear is measured by averaging over large numbers of galaxies pairs. For the result to be meaningful, sources of bias caused by systematic effects need to be sub-dominant.
Systematic effects can be mitigated through instrument design, but some need to be modelled and removed from the data. In order to determine how such systematic effects can bias the cosmic shear measurements -- and cosmological parameter inference -- a series of papers derived analytic expressions that represented the 
measurement and modelling processes involved. Following an initial study by \cite{Paulin-Henriksson08} that focused on point spread function (PSF) requirements, \citet[][M13 hereafter]{Massey13} presented a more general analytic framework that captures how various systematic effects affect the measurements of galaxy shapes. This study provided the basis for a detailed breakdown of systematic effects for {\it Euclid} by \citet[][C13 hereafter]{Cropper13}, which has been used in turn to derive requirements on the performance of algorithms and supporting data. Another approach, based on Monte Carlo Control Loops (MCCL), has also been presented \citep{Bruderer2018,2014PDU.....3....1R} where one uses a forward modelling approach to calibrate the shear measurement.

Although these previous studies provide a convenient way to compare the impact of various sources of bias, their analytic 
nature means that particular assumptions are made, and they cannot capture the full realism of a cosmic shear survey.
Therefore we revisit the issue in this paper for a number of reasons:
\begin{enumerate}
\item
In order to avoid an implicit preference for 
implementation, the derivations in \citetalias{Massey13} are scale-independent i.e. they do not depend on angle $\theta$ or multipole $\ell$ explicitly. In more realistic scenarios, such as the ones we consider here, spurious signals are introduced on specific spatial and angular scales on the celestial sphere. For example, 
the PSF model is determined from the full instrument field-of-view, whereas detector effects, such as charge transfer inefficiency (CTI) occur on the scale of the region served by a single readout register on a CCD. In addition, the biases may depend on observing strategy or time since launch. This is particularly true for CTI, 
which is exacerbated by radiation damage, and thus increases with time \citep{10.1093/mnras/stu012, Israel15}. 
An initial study of the implications of scale-dependent scenarios was presented 
in \cite{Kitching16} who found that survey strategy can play a critical role in the case of time-dependent effects. Their results suggest the expected biases in cosmological parameters  may be reduced if the correct scale dependencies are considered. 
\item
The residual systematic effects may depend on the region of the sky that is observed. For example, the model of the PSF can 
be constrained to a higher precision when the density of stars is higher. 
On the other, hand they may also have an adverse effect on the galaxy shape measurement of the shear \citep{Hoekstra17}. 
The impact of CTI depends on the sky background level, and thus is a function of ecliptic latitude, whereas Galactic extinction may introduce biases in the determination of photometric redshift that depend on Galactic latitude (and longitude). These subtle variations across the survey should be properly accounted for, and their 
impact on the main science objectives of {\it Euclid} evaluated. 
\item 
In the analytic results of e.g., \citetalias{Cropper13}, a distinction was made between `convolutive' (caused by PSF) and `non-convolutive' contributions. 
The impact of the former, such as the PSF, are relatively easy to propagate, because it is 
typically clear how they depend on galaxy properties. The latter, however, which include 
biases introduced by CTI, are more complicated to capture, because their dependence on galaxy properties such as 
size and flux can be non-linear. Moreover, the allocations implicitly assume that residual errors are 
independent, because correlations between effects could not be easily included. Hence, the impact of a 
more realistic error propagation needs to be examined. 
\item
The interpretation of the requirements presented in \citetalias{Cropper13} is unclear, in particular whether they should be considered as values that are never to be exceeded, the mean of a distribution of possible biases, or upper limits corresponding to a certain confidence limit. As shown below, we expect our limited knowledge of the system to result in probability density distributions of biases that should be consistently combined to evaluate the overall performance.
\item 
Finally, in \cite{Kitching19} we show that these previous studies made simplifying assumptions with regard to the analytic relationship between position-dependent biases and the cosmic shear statistics, where the correct expression involves second and third order terms. This motivates our study in two ways. Firstly the correct expression involves previously unstudied terms. Secondly, the correct expression is computationally demanding, meaning its calculation is intractable for realistic cosmic shear measurements.
\end{enumerate}
In this paper we present a general framework for investigating systematic effects that addresses all these issues, but does not require full image-level end-to-end simulations (which would require fully realistic mock data and data processing stages). Instead our approach starts at the object catalogue level, and  systematic effects are propagated through a chain of processes on an object-by-object basis. This does not mean that systematic effects are not in common between galaxies, but it assumes that the measurement process is. This is a reasonable assumption for weak lensing studies where the shape measurement itself is confined to a narrow angular region about the vicinity of the galaxy on the sky. This allows us to create scenarios where systematic effects are calculated in a more realistic fashion, starting from a catalogue of sources with appropriate parameters, and  propagated all the way to the evaluation of cosmological parameters. 
Although this approach may not capture all correlations between systematic effects (this can only be achieved through a full end-to-end simulation of the pipeline), it does present a major advance over the initial studies presented in \citetalias{Massey13} and \citetalias{Cropper13}.

We describe the general framework in more detail in Sect. \ref{sec:framework}, where we also discuss the properties of the input catalogue, sky parameters and observational characteristics. Results are presented in Sect. \ref{e2e}. A more complete exploration of the many possible sources of bias for {\it Euclid} is deferred to future work, but in Sect. \ref{sec:casestudies} we consider a few case studies: in Sect. \ref{sec:psf} the residuals in the PSF correction, and in Sect. \ref{sec:cti} the impact of imperfections in the correction for CTI. Although the performance analysis in this paper takes {\it Euclid} as a reference mission, the framework is sufficiently general that it can be applied to any future Stage IV weak gravitational lensing survey \citep[e.g.][]{LSST,WFIRST}. 
\begin{figure*}
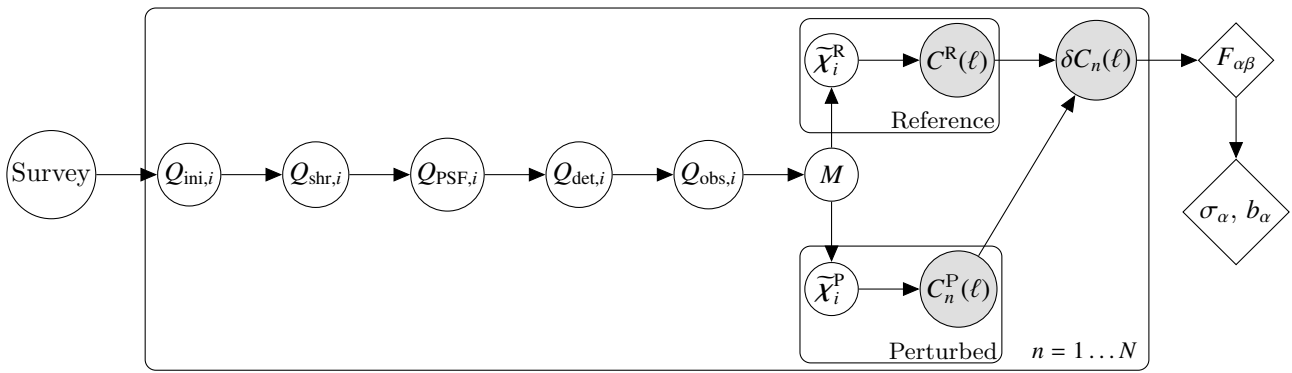

  \centering
  \tikz{ %
    \node[latent] (sur) {Survey} ; %
    \node[latent, right=of sur] (g) {$Q_{{\rm ini},i}$} ; %
    \node[latent, right=of g] (eI) {$Q_{{\rm shr},i}$} ; %
    \node[latent, right=of eI] (eP) {$Q_{{\rm PSF},i}$} ; %
    \node[latent, right=of eP] (eD) {$Q_{{\rm det},i}$} ; %
    \node[latent, right=of eD] (eO) {$Q_{{\rm obs},i}$} ; %
    \node[latent, right=of eO] (M) {$M$} ; %
    
    \node[latent, above=of M] (chiR) {$\widetilde {\bm \chi}^{\rm R}_i$} ; %
    \node[obs, right=of chiR] (CR) {$C^{\rm R}(\ell)$} ; %

    \node[latent, below=of M] (chiP) {$\widetilde {\bm \chi}^{\rm P}_i$} ; %
    \node[obs, right=of chiP] (CP) {$C^{\textrm P}_{n}(\ell)$} ; %

    \node[obs, right=of CR] (dC) {$\delta C_n(\ell)$} ; %

    \node[det, right=of dC] (F) {$F_{\alpha\beta}$} ; %
    \node[det, below=of F] (E) {$\sigma_{\alpha}$, $b_{\alpha}$} ;

    \platet[inner sep=0.15cm, xshift=-0cm, yshift=0cm] {plate2} {(g) (eI) (eP) (eD) (M) (chiR) (CR) (CP) (dC) (chiP)} {$n=1\dots N$}; %
    \platet[inner sep=0.05cm, xshift=-0cm, yshift=0cm] {plate1} {(chiP) (CP)} {Perturbed}; %
    \platet[inner sep=0.05cm, xshift=-0cm, yshift=0cm] {plate2} {(chiR) (CR)} {Reference}; %

    \edge {sur} {g} ; %
    \edge {g} {eI} ; %
    \edge {eI} {eP} ; %
    \edge {eP} {eD} ; %
    \edge {eD} {eO} ; %
    \edge {eO} {M} ;
    \edge {M} {chiR} ;
    \edge {chiR} {CR} ;
    \edge {CR} {dC} ;

    \edge {M} {chiP} ;
    \edge {chiP} {CP} ;
    \edge {CP} {dC} ;

    \edge {dC} {F} ; 
    \edge {F} {E} ;
}
\caption{The overall structure of the concept as described in the main text. The quadrupole moments $Q$ are initiated with intrinsic moments, and then modified by incorporating the shear, PSF and detector effects. Survey characteristics such as dither pattern, slew pattern and observation time are entered in the initial catalogue. Then a measurement process $M$ converts the observed moments to polarisations. The estimation of the galaxy polarisation is then made (as described in Eqs.~\ref{ref} and \ref{pert}). This is done per object. Then a power spectrum for the reference and the perturbed scenarios is computed. For the perturbed line the PSF and detector moments are drawn from distributions that represent the measurement uncertainty as described in the text. This process is repeated for 150 random realisations for the set of galaxies that are in the input catalogue. Finally the residual power spectrum is computed per realisation, and the statistics of each of the realisations is passed onto the Fisher matrix, from which uncertainties and biases of dark energy parameters are calculated. White circles indicate moment space, where modifications are performed on an object-by-object basis. Gray circles indicate ensemble average in the harmonic space. Diamonds show cosmological parameter space.
}
\label{Bnet}
\end{figure*}

\section{General framework}
\label{sec:framework}
The general framework we present is a causally connected pipeline, or transfer function-like methodology. This pipeline modifies the values of quantities associated with each individual galaxy according to the effects that the instrument and measurement processes have. These are in turn used to compute cosmic shear power spectra to evaluate the impact on cosmological parameter inference. The general framework is captured in Fig. \ref{Bnet}, that we summarise in Sect. \ref{pipesum}.

\subsection{Causally connected pipeline}
As light propagates from a galaxy, several processes occur that act to transform a galaxy image. We represent this  as a series of sequential processes, or a pipeline, that are causally ordered, for example 
\bee
I^+_{{\rm ini},i} \rightarrow I^+_{{\rm shr},i}(I^+_{{\rm ini},i}) \rightarrow
I^+_{{\rm PSF},i}(I^+_{{\rm shr},i}) \rightarrow
I^+_{{\rm det},i}(I^+_{{\rm PSF},i}) \rightarrow
M(I^+_{{\rm det},i}) \;,
\label{pipeline}
\eee
where $I$ is a surface brightness, the subscripts $i$ refer to an object (a galaxy in our case), the other subscripts refer to the addition of an effect (labelled as a $+$): where in this example `shr' labels shear, `PSF' labels the PSF, det labels the detector, etc. The pipeline is initiated by a projected initial (intrinsic) surface brightness distribution $I^+_{{\rm ini},i}$ for object $i$ that is modified/transformed via a series of processes -- the shearing by large-scale structure, the convolution by the PSF -- that depend on the preceding step. The last step $M$ represents a measurement process that converts the observed surface brightness distribution into quantities that can be used for science analyses. 
Eq.~(\ref{pipeline}) is an example, that includes shear and PSF effects, of a more general framework that we define here
\bee 
I^+_{{\rm ini},i} \rightarrow I^+_{\alpha,i}(I^+_{{\rm ini},i})\rightarrow
I^+_{\alpha+1,i}(I^+_{\alpha,i})\rightarrow \dots \rightarrow M(I^+_{\alpha+n,i}) \;,
\eee
where $\alpha$ is some general process that modifies the surface brightness distribution of object $i$ that precedes process $\alpha+1$, and so forth. In this paper we focus only on the impact of PSF and detector effects on cosmic shear analyses, but emphasise that the approach is much more general. It can be readily extended to include more effects, such as photometric errors, spectral energy distribution (SED) dependent effects, or the impact of masking. These will be explored in future work.  

The objects in question for weak lensing measurements are stars -- which are used for PSF determination -- and galaxies. The primary quantities of interest for these galaxies are the quadrupole moments of their images, which can be combined to estimate polarisations and sizes. The unweighted quadrupole moments $Q_{i,mn}$ of a projected surface brightness distribution (or image)  $I_i({\bm x})$ are defined as
\begin{equation}
\label{Qeq}
Q_{i,mn}=\frac{1}{F}\int \d^2{\bm x}\, x_m \, x_n \,I_i({\bm x})\;,
\end{equation}
where $F$ is the total observed flux, $m$ and $n$ are $(1,2)$ corresponding to orthogonal directions in the image plane, and we assumed that the image is centred on the location where
the unweighted dipole moments vanish. We can combine the quadrupole moments to obtain an estimate of the size $R=\sqrt{Q_{11}+Q_{22}}$, and shape of a galaxy through the complex polarisation,
or third eccentricity\footnote{We note that this is the same combination of moments used by \citetalias{Massey13}, but who refer to the polarisation by a different name `ellipticity' denoted as $\epsilon$.}
\begin{equation}
\label{chiQ}
{\bm \chi}=\frac{Q_{11}-Q_{22}+2 {\rm i}Q_{12}}{Q_{11}+Q_{22}}.
\end{equation}
Therefore, the pipeline process for the cosmic shear case is similar to the one given by Eq.\,(\ref{pipeline}), but for the quadrupole moments of the surface brightness distribution. In this case each process acting on the surface brightness distribution is replaced by its equivalent process acting on the quadrupole distribution; and the final measurement process is the conversion of quadrupole moments into polarisation:
\begin{align}
Q^+_{{\rm ini},i}&\rightarrow Q^+_{{\rm shr},i}(Q^+_{{\rm ini},i})\rightarrow Q^+_{{\rm PSF},i}(Q^+_{{\rm shr},i})\nn
&\rightarrow Q^+_{{\rm det},i}(Q^+_{{\rm PSF},i})\rightarrow 
{\bm\chi}_{\rm obs,i}(Q^+_{{\rm det},i})\;,\label{Qpipeline}
\end{align}
where we suppress the $mn$ subscripts for clarity. In this expression ${\bm\chi}_{\rm obs,i}$ is the observed polarisation for object $i$ that is a function of $Q^+_{{\rm det},i}$, where these quantities are related by Eq.~(\ref{chiQ}) in the general case. The result is then used for cosmic shear analysis.  
Importantly, at each stage in the pipeline, the relevant quantities that encode the intrinsic ellipticity/shear/PSF/detector effects, instead of being fixed for all objects, can be drawn from distributions or functions that capture the potential variation owing to noise in the system and the natural variation of object and instrumental properties. 


\subsection{Reference and perturbed scenarios}
Next we introduce the concept of a \emph{reference} scenario,
representing the ideal case, and a \emph{perturbed} scenario that results in biased estimates caused by misestimation and uncertainty in the inferred values of the quantities that are included in the set of causally linked processes as described in Eq.\,(\ref{pipeline}). We define these below. 
\begin{itemize}
\item[] 
{\bf Reference}: In this scenario the systematic effects that have been included in the pipeline are perfectly known, so that in the final measurement process their impact can be fully accounted for and reversed. In this case the distribution of parameter values that are used to undo the biases are all delta-functions centred on the reference values, i.e. there is no uncertainty in the system.  
\item[] 
{\bf Perturbed}: In this scenario systematic effects that have been included in the pipeline are not known perfectly. As a consequence the corrections result in biased measurements. In this case relevant quantities that are used to undo the systematic effects are drawn from probability distributions that represent the expected level of uncertainty. 
\end{itemize}
We can then define the elements in a pipeline for each scenario. The difference between the observed reference polarisation for a given object, and the observed perturbed polarisation is a realisation of expected polarisation uncertainty caused by a semi-realistic treatment of systematic effects in a data reduction scenario. We explain this further using the specific example with which we are concerned in this paper: the assessment of cosmic shear performance. 

In our case, the output of the pipeline process, Eq.~(\ref{Qpipeline}), leads to a set of measured polarisations and sizes, that represent the true response of the system i.e. an ellipticity catalogue that includes the cumulative effects of the individual processes as they would have occurred in the real instrument and survey. As detailed in \citetalias{Massey13}, we can compute how PSF and detector effects change the polarisation and size of a galaxy\footnote{We note that this formalism does not capture non-linear effects whereby the change in moments caused by PSF or detector effects may depend on a galaxy's intrinsic shape and brightness. We leave a relaxation of this linearity assumption to future work.}:
\begin{align}
{\bm\chi}_{{\rm obs},i}&={\bm\chi}_{{\rm ini},i}+{\bm\chi}_{{\rm shr},i}\nn
&+\left(\frac{R^2_{{\rm PSF},i}}{R^2_{{\rm PSF},i}+R^2_{{\rm ini},i}+R^2_{{\rm shr},i}}\right)\,
\left({\bm\chi}_{{\rm PSF},i}-
{\bm\chi}_{{\rm ini},i}-{\bm\chi}_{{\rm shr},i}\right)\nn
&+{\bm\chi}_{{\rm det},i}\;,
\end{align}
where ${\bm\chi}_{{\rm obs},i}$ is the observed polarisation, ${\bm\chi}_{{\rm ini},i}$ is the intrinsic/unlensed polarisation, ${\bm\chi}_{{\rm shr},i}$ is the induced polarisation caused by the applied shear $\bm\gamma$, ${\bm\chi}_{{\rm PSF},i}$ is the polarisation of the PSF, and ${\bm\chi}_{{\rm det},i}$ is the detector-induced polarisation; the same subscripts apply to the $R^2$ terms ($R=\sqrt{Q_{11}+Q_{22}}$, see Eq. \ref{Qeq}). The relation between the applied shear, $\bm\gamma$, and the corresponding change in polarisation, $\bm\chi_{\rm shr}$, is quantified by the shear polarisability ${\textbf{\textsf{P}}}^\gamma$ so that 
\begin{equation}
\label{Pgamma}
    {\bm\chi}_{\rm shr}={\textbf{\textsf{P}}}^\gamma {\bm\gamma}
\end{equation} 
\citep{KSB95}. The shear polarisability depends on the galaxy morphology,
but it can be approximated by the identity tensor times a real scalar ${\textbf{\textsf{P}}}^\gamma=(2-\langle\chi_{\rm ini}^2\rangle){\textbf{\textsf{I}}}$ (where ${\textbf{\textsf{I}}}$ is the identity matrix) in the case of unweighted moments \citep{Rhodes00}. We simplify this equation, in terms of notation, to
\bee 
\label{simplem}
{\bm\chi}_{{\rm obs},i}={\bm\chi}_{{\rm gal},i}+
f_i\,({\bm\chi}_{{\rm PSF},i}-{\bm\chi}_{{\rm gal},i})+
{\bm\chi}_{{\rm det},i}\;,
\eee
where ${\bm\chi}_{{\rm gal},i}={\bm\chi}_{{\rm ini},i}+{\bm\chi}_{{\rm shr},i}$ (the polarisation that would be observed given no PSF or detector effects), and
\bee
f_i=\frac{R^2_{{\rm PSF},i}}{R^2_{{\rm obs},i}}\;.
\eee
These quantities are constructed from the corresponding quadrupole moments in Eq.~(\ref{Qpipeline}). 

Given a set of observed galaxy polarisations and sizes and perfect knowledge of the systematic effects Eq.\,(\ref{simplem}) can be inverted, yielding an estimate for the galaxy shape in the {\bf reference} case given by
\bee
\label{ref}
\widetilde{\bm\chi}^{\rm R}_{{\rm gal},i}=\frac{{\bm\chi}_{{\rm obs},i}-f_i^{\rm R}{\bm\chi}^{\rm R}_{{\rm PSF},i}-{\bm\chi}^{\rm R}_{{\rm det},i}}{1-f_i^{\rm R}}\;, 
\eee
where the superscript $R$ denotes the reference case. 
In this case the quantities ${\bm\chi}^{\rm R}_{{\rm PSF},i}$, ${\bm\chi}^{\rm R}_{{\rm det},i}$, and $f_i^{\rm R}$ are known exactly and constructed from the quadrupole moments in Eq.\,(\ref{Qpipeline}), and we obtain (trivially) the underlying true $\widetilde{\bm\chi}^{\rm R}_{{\rm gal},i}={\bm\chi}_{{\rm gal},i}$. Even though this is a trivial inversion we nevertheless perform this step since in general the measurement process may not be exactly invertable. 

In the {\bf perturbed} case, the uncertainties in the measurement and modelling process result in a set of estimated values that include residual effects of the PSF and detector
\bee
\label{pert}
\widetilde{\bm\chi}^{\rm P}_{{\rm gal},i}=\frac{{\bm\chi}_{{\rm obs},i}-f_i^{\rm P}{\bm\chi}^{\rm P}_{{\rm PSF},i}-{\bm\chi}^{\rm P}_{{\rm det},i}}{1-f_i^{\rm P}}\;,
\eee
where the superscript $P$ denotes the perturbed case. Here ${\bm\chi}^{\rm P}_{{\rm PSF},i}$, ${\bm\chi}^{\rm P}_{{\rm det},i}$, and $f_i^{\rm P}$ are constructed from the quadrupole moments drawn from relevant probability distributions that represent uncertainties in the system. The resulting polarisation estimates correspond to a realisation of the system that encodes the expected uncertainty in our understanding of PSF and detector effects. Each of these steps is then repeated for realisations of the probability distributions present in the perturbed quantities. The implementation of these probability distributions for the PSF and CTI cases are detailed in Appendices~A and~B. 

To convert the estimated reference and perturbed polarisations to their corresponding shear estimates we use
\begin{equation}
\label{Pgamma2}
    \tilde\gamma_{\rm gal} = [{\textbf{\textsf{P}}^\gamma}]^{-1}{\bm \chi}_{\rm gal},
\end{equation}
that provides a noisy, but unbiased estimate of the shear $\bm\gamma$ \citepalias{Massey13}. We note that ${\textbf{\textsf{P}}}^\gamma$ does not change between the reference and perturbed cases. In practice, shape measurement algorithms use weighted moments to suppress the noise in the images, which changes the shear polarisation compared to the unweighted case. The correction for the change in shape caused by the weight function depends on the higher-order moments of the surface brightness \citep{Melchior11} and is a source of shape measurement bias that can be quantified using image simulations \citep[e.g.][]{Hoekstra17}. This also leads to a sensitivity to spatial variations in the colours of galaxies if the PSF is chromatic \citep{Semboloni13,Er18}. However, for the study presented here, this complication can be ignored as we implicitly assume that the biases in the shape measurement algorithm have been accounted for to the required level of accuracy \citepalias{Cropper13}.

In future work, we will include more effects in the perturbed scenario. Observable quantities $\widetilde{\bm\chi}^{\rm P}_i$ can be generalised to a function of redshift and wavelength, i.e. $\widetilde{\bm\chi}^{\rm P}_{{\rm gal},i}(z,\lambda)$. We will then explore the effects of masking, shape measurement errors, photometric errors, and SED variations within a galaxy.

\subsection{Shear power spectrum estimation}
The estimated polarisations contain 
\begin{align}
\widetilde{\bm\chi}^{\rm R}_{{\rm gal},i}&={\bm\chi}_{{\rm gal},i},\nn
\widetilde{\bm\chi}^{\rm P}_{{\rm gal},i}&\approx{\bm\chi}_{{\rm gal},i}+{\bm\chi}^{\delta}_{{\rm gal},i},\label{eq:chiRP}
\end{align}
where ${\bm\chi}^{\delta}_{{\rm gal},i}$ is the change in polarisation. We assume higher-order terms are subdominant, i.e.\ terms involving $({\bm\chi}^{\rm P}_{{\rm gal},i})^n\approx 0$ for $n>1$. ${\bm\chi}^{\delta}_{{\rm gal},i}$ is caused by the uncertainty in systematic effects, that is defined by expanding the denominator in Eq.~(\ref{pert}) to linear order, and substituting Eq.~(\ref{simplem}):
\begin{equation}
    \widetilde{\bm\chi}^{\rm P}_{{\rm gal},i}\approx {\bm\chi}_{{\rm gal},i}+[f^{\rm R}_i\,({\bm\chi}^{\rm R}_{{\rm PSF},i}-{\bm\chi}^{\rm R}_{{\rm gal},i})+{\bm\chi}^{\rm R}_{{\rm det},i}-{\bm\chi}^{\rm P}_{{\rm PSF},i}-(1/f_i^{\rm P}){\bm\chi}^{\rm P}_{{\rm det},i}]\;,
\end{equation}
where the denominator in Eq.~(\ref{pert}) is expanded by assuming $f^{\rm R}_i\ll 1$. 

The polarisations in Eq.~(\ref{eq:chiRP}) can be converted to estimates of the corresponding shears using Eq.~(\ref{Pgamma}) and Eq.~(\ref{Pgamma2}), $\widetilde{\bm\gamma}^{\rm R}$ and $\widetilde{\bm\gamma}^{\rm P}$. 
These can be subsequently used to calculate shear power spectra, and the residual between the reference and perturbed spectra: 
\begin{align}
\delta C_n(\ell)&=C^{\rm P}_n(\ell)-C^{\rm R}(\ell)\nonumber\\
&\approx C^{{\rm gal}-\delta}_n(\ell)+C^{ \delta-{\rm gal}}_n(\ell)+C^{\delta-\delta}_n(\ell)\;,
\end{align}
where 
\bee
C^{\rm P}_n(\ell)=\frac{1}{2\ell+1}\sum_{m=-\ell}^{\ell}
\widetilde{\bm\gamma}^{\rm P}_{\ell m}(\widetilde{\bm\gamma}^{\rm P}_{\ell m})^*\;,
\eee
where $\widetilde{\bm\gamma}^{\rm P}_{\ell m}$ are the spherical harmonic coefficients of the perturbed shear field i.e.
\bee 
\widetilde{\bm\gamma}^{\rm P}_{\ell m}=\sqrt{\frac{1}{2\pi}}\sum_{i}
\widetilde{\bm\gamma}^{\rm P}_i\;_2Y_{\ell m}(\theta_i,\phi_i)\;.
\eee
In the above expressions $(\theta_i,\phi_i)$ is the angular coordinate of galaxy $i$, the ${}_2Y_{\ell m}(\theta_i,\phi_i)$ are the spin-weighted spherical harmonic functions, and a $^*$ refers to a complex conjugate. Similarly for the reference case $C^{\rm R}_n(\ell)$. $C^{\rm P}_{n}(\ell)$ is a realisation $n$ of one that may be observed given the limited knowledge of uncertainties in systematic effects. We can split the residual power spectrum into three terms; $\delta-\delta$ quantifies the auto-correlation of the systematic uncertainties; gal-$\delta$ and $\delta$-gal are the cross-correlation power spectra between the systematic uncertainties and the true cosmological signal (i.e. the one that would have been observed if all systematic effects were perfectly accounted for).

Although selection effects can result in a correlation between the shear and systematic effects, we stress that we are interested in residual effects, and thus implicitly assume that such selection effects have been adequately accounted for. Hence, when taking an ensemble average over many realisations, we are left with $\langle\delta C_n(\ell)\rangle=C^{\delta-\delta}(\ell)$ as the mean of these additional terms should reduce to zero and any variation is captured in the error distribution of the $\delta C(\ell)$'s. Hence we can determine the power spectrum caused by uncertainties in systematic effects.

We sample from all parameter probability distributions in the perturbed case, and compute the mean and variance over the resulting ensemble of $\{\delta C_n(\ell)\}$. In the cases where random numbers are required for the reference case, care must be taken to ensure that the seed is the same in the reference and perturbed cases.

\subsection{Comparison to previous work}
To compare to previous work, in \citetalias{Massey13} generic non-parametric realisations of $\delta C(\ell)$ were generated and used to place conservative limits on a multiplicative and additive fit to such realisations $\delta C(\ell) ={\cal M}\,C^{\rm R}(\ell)+{\cal A}$, where ${\cal M}$ and ${\cal A}$ are constant so that biases in the dark energy parameters, using Fisher matrix predictions, were below an acceptable value. This represents a worst case, because the residual power spectra are assumed to be proportional to the cosmological signal (apart from the additive offset). In \cite{Kitching16}, simple models for systematic effects were used to create simplified but realistic $\delta C(\ell)$ values. In \cite{Taylor18}, the constant multiplicative and additive formulation was generalised to include the propagation of real-space multiplicative effects into power spectra as a convolution. 
In \cite{Kitching19} the full expression for the analytic propagation of constant and scale-dependent multiplicative and additive biases is derived. This reveals that the analytic propagation of biases into cosmic shear power spectra involves second- and third-order terms that result in an intractable calculation for high-$\ell$ modes.
Our approach, therefore, differs from the earlier works in that it captures any general scale and redshift dependence on an object-by-object level, and, very importantly, creates $\delta C(\ell)$ values that correctly incorporate the uncertainty in the system. This procedure enables a complete evaluation of the performance, that differs from a true end-to-end evaluation only in that we do not use the images and image-analysis algorithms that will be used to analyse the real data.  

These catalogue-level simulations have the major advantage that they are much faster than full end-to-end image simulations, allowing for realisations of systematic effects to be computed so that a full probability distribution of the effect on the cosmological performance of the experiment can be determined. This allows us to explore various survey strategies and other trade-off considerations, whilst capturing most of the complexities of the full image-based analysis. The catalogue-level simulations include survey-specific features, such as the detector layout, survey tiling and PSF pattern (see \S\ref{sec:survey}). It also allows for foreground sky models to be included to account for variations in Galactic extinction, star density and Zodiacal background. Calibration uncertainties can be incorporated by adjusting the probability density distributions of the relevant parameters accordingly.

\subsection{Propagation to cosmological parameter estimation}
\label{cosmo}
To assess the impact of the power spectrum residuals on cosmological parameter inference, we use the Fisher matrix, and bias formalism \citep{Kitching08,2008MNRAS.391..228A,Taylor18}. Here we provide a short summary of the Fisher matrix formalism used in those papers.  

In general, a change in the power spectrum caused by a residual systematic effect can influence the size of the confidence region about any parameter as well as the maximum  likelihood location. In this paper we only consider the change in the maximum likelihood position.

The expected confidence regions for the cosmological parameters can be expressed using the Fisher matrix, which is given by 
\bee
F_{\alpha\beta}=\sum_{jk,\ell}{\mathcal F}_{jk}(\ell)\;
\;\frac{\partial C_{jk}(\ell)}{\partial \alpha}\;\frac{\partial C_{jk}(\ell)}{\partial \beta}\;,
\eee
where $m, n$ are redshift bin pairs and the Greek letters denote cosmological parameter pairs. ${\mathcal F}_{mn}(\ell)$ is given by
\citep{Hu99} 
\bee
{\mathcal F}_{jk}(\ell)=\frac{f_{\rm{sky}}\,(2\ell+1) \,}{2[C_{jk}(\ell)+N_{jk}(\ell)]^2}\;,
\eee
where $f_{\rm{sky}}$ is the fraction of the sky observed and $C_{jk}$ the cross power spectrum between redshift slices $j$ and $k$. The noise power spectrum is defined as $N_{jk}(\ell)=\sigma_{\chi_{\rm ini}}^2\delta_{jk}/N_{{\rm g},j}$, where $N_{{\rm g},j}$ is the total number of galaxies in bin $j$ for full sky observation and $\delta_{jk}$ is a Kronecker delta. The intrinsic shape noise is quantified by $\sigma_{\chi_{\rm ini}} = 0.3$, the dispersion per ellipticity component. The signal power spectrum in the denominator is $C^{\rm R}_{jk}(\ell)$ in the reference case and $C^{\rm P}_{jk}(\ell)$ for each realisation $n$ of the perturbed case. This can be used to compute the expected marginalised, cosmological parameter uncertainties $\sigma_{\alpha}=[(F^{-1})_{\alpha\alpha}]^{1/2}$.

The changes in the maximum likelihood locations  of the cosmology parameters (i.e. biases) caused by a change in the power spectrum can also be computed for parameter $\alpha$ as
\bee
\label{eq:bias}
b_{\alpha}=-\sum_{\beta}(F^{-1})_{\alpha\beta}\;B_{\beta}\;,
\eee 
where the vector $B$ for each parameter $\beta$ is given by 
\bee
B_{\beta}=\sum_{jk,\ell}{\mathcal F}_{jk}(\ell)\;\delta C_{jk}(\ell)\;\frac{\partial C_{jk}(\ell)}{\partial \beta}\;.
\eee
We note that the biases computed here are the one-parameter, marginalised biases and that this may result in optimistic assessments for multi-dimensional parameter constraints. For a multi-dimensional constraint may be biased by more than $1$-sigma along a particular degenerate direction, and yet the marginalised biases may both be less than $1$-sigma.  

The fiducial cosmology we have used in the Fisher and bias calculations is a flat $w_0w_a$CDM cosmology with a redshift-dependent dark energy equation of state, defined by the set of parameters 
$\Omega_{\rm m}$, $\Omega_{\rm b}$, $\sigma_8$, $w_0$, $w_a$, $h$, $n_s$; these are 
the matter density parameter; baryon density parameter; the amplitude of matter fluctuations on $8h^{-1}$Mpc scales -- a normalisation of the power spectrum of matter perturbations; the dark energy equation of state parameterised by $w(z)=w_0+w_a z/(1+z)$; the Hubble parameter $H_0=100h\,{\rm km\,s^{-1}\,Mpc^{-1}}$; and
the scalar spectral index of initial matter perturbations, respectively. The fiducial values are taken from the 
\emph{Planck} maximum likelihood values \citep{Planck13_paperXVI}. The uncertainties and biases we quote on individual dark energy parameters are marginalised over all other parameters in this set. The survey characteristics we use are based on a \emph{Euclid}-like wide survey \citep{Laureijs11} that has an area of
$15$\,$000$ deg$^2$, a median redshift of
$z_{\rm med}=0.9$, and a galaxy number density of $30\,{\rm arcmin}^{-2}$. Throughout we use an $\ell$ range $10\leq\ell\leq 4000$. We use the weak lensing only Fisher matrix from the \emph{Euclid} Inter Science Taskforce (IST) forecasting paper \citep{EC}, where further details can be found; for a flat $w_0w_a$CDM cosmology the marginalised $1$-sigma errors from that paper (Table 11) are: $\sigma(\Omega_{\rm m})=0.034$, 
$\sigma(\Omega_{\rm b})=0.42$, 
$\sigma(w_0)=0.14$, 
$\sigma(w_a)=0.48$, 
$\sigma(h)=0.20$, 
$\sigma(n_s)=0.030$, 
$\sigma(\sigma_8)=0.013$ for an optimistic setting (defined in that paper).

In this paper we will only quote biases on dark energy parameters, relative to the expected parameter uncertainty. Finally, we note that we do not consider redshift-dependent systematic effects. Consequently, $\delta C_{mn}(\ell)=\delta C(\ell)\delta_{mn}$, i.e. the systematic effects are equal for each tomographic redshift bin.

\subsection{Summary of the pipeline}
\label{pipesum}
In Fig.~\ref{Bnet} we summarise the overall architecture of the current concept. This propagates the changes in the quadrupole moments, converts these to observed polarisation, determines the estimated galaxy polarisation, and then power spectra and the residuals. The steps are listed below.

\begin{itemize}
    \item Survey: specifies input positional data for each galaxy, for example the position, dither pattern, slew pattern, observation time, etc.;
    \item $Q_{{\rm ini},i}$: initial, intrinsic quadrupole moments are assigned to a galaxy; 
    \item $Q_{{\rm shr},i}$: shear effects are included for each galaxy in the form of additional quadrupole moments;
    \item $Q_{{\rm PSF},i}$: PSF effects are included for each galaxy, these can be drawn from a distribution representing the variation in the system;
    \item $Q_{{\rm det},i}$: detector effects are included for each galaxy, these can be drawn from a distribution representing the variation in the system;
    \item $Q_{{\rm obs},i}$: observational effects are included for each galaxy such as the impact of shape measurement processes. In this paper these are not included, but we include them in the pipeline for completeness; 
    \item $M$: moment measurements are converted into polarisations $\chi_{{\rm obs},i}$. At this step, where the systematic effects are removed, the reference and perturbed lines separate;
    \item $\widetilde {\bm\chi}^{\rm R}_i$: a reference polarisation is computed, from Eq.~(\ref{ref}), which includes ${\bm\chi}^{\rm R}_{{\rm PSF},i}$, ${\bm\chi}^{\rm R}_{{\rm det},i}$, and $f_i^{\rm R}$ that are the same values used in the construction of $\chi_{{\rm obs},i}$; 
    \item $\widetilde {\bm\chi}^{\rm P}_i$: a perturbed polarisation is computed, from Eq.~(\ref{pert}), which includes ${\bm\chi}^{\rm P}_{{\rm PSF},i}$, ${\bm\chi}^{\rm P}_{{\rm det},i}$, and $f_i^{\rm P}$ constructed from  quadrupole moments drawn from relevant probability distributions that represent uncertainties in the system; 
    \item $C^{\rm R}(\ell)$: computes the power spectrum of $\widetilde {\bm \chi}^{\rm R}_i$; 
    \item $C^{\rm P}(\ell)$: computes the power spectrum of $\widetilde {\bm \chi}^{\rm P}_i$; 
    \item $\delta C(\ell)$: computes the residual power spectrum for realisation $n$;
    \item $F_{\alpha\beta}$: computes the Fisher matrix and biases given the perturbed power spectrum that can be used to derive uncertainties $\sigma_{\alpha}$ and biases $b_{\alpha}$.
\end{itemize}

\section{End-to-end pipeline}
\label{e2e}
Having introduced the general formalism, we now describe the details of the current pipeline. As we work at the catalogue-level, we have full flexibility over the steps that are included in or excluded from the pipeline. Furthermore, the approach (and code) is modular, giving us full flexibility in terms of developing the pipeline further. As certain steps in the pipeline mature, the relevant modules can be updated with increasingly realistic performance estimate. 

\subsection{Input catalogue}
\label{sec:inputcatalog}

To evaluate the performance we need an input catalogue that contains galaxies with a range of sizes, magnitudes, and redshifts\footnote{These properties are not used in the tomographic bin definition used in the Fisher matrix calculation, which is a sophistication that will be included in later iterations of the pipeline.}. It is also important that the
catalogue captures spatial correlations in galaxy properties, e.g., clustering, because the morphology and SED of a galaxy correlate with its local environment.

\subsubsection{Mock catalogue: MICE}

Here we use the Marenostrum Institut de Ci{\`e}ncies de l'Espai (MICE) Simulations catalogue to assign galaxy properties, such as magnitude, RA, dec, shear and etc.. It is based on the DES-MICE catalogue and designed for {\it Euclid} \citep{Fosalba15a,Fosalba15b, Crocce15}. It has approximately 19.5 million galaxies over a total area of 500 deg$^2$, with a maximum redshift of $z\simeq 1.4$. The catalogue is generated using a Halo Occupation Distribution (HOD) to populate Friends of Friends (FOF) dark matter halos from the MICE simulations \citep{Carretero15}. The catalogue has the following observational constraints: the luminosity function is taken from \cite{Blanton03}; the galaxy clustering as a function of the luminosity and colour follows \cite{Zehavi11}; and the colour-colour distributions are taken from COSMOS \citep{Scoville07}.

A model for galaxy evolution is included in MICE to mimic correctly the luminosity function at high redshift. The photometric redshift for each galaxy is computed using a photo-$z$ template-based code, using only Dark Energy Survey (DES) photometry; see \cite{Fosalba15a,Fosalba15b, Crocce15} for details of the code. Our magnitude cut is placed at $20.0 \le m_{\rm VIS} \le 25.0$ in the {\it Euclid} VIS band. We use a $10\times 10$ deg$^2$ area of the catalogue, containing approximately 4 million galaxies.

\subsubsection{Intrinsic polarisations}
The MICE catalogues contain the information about the position, redshift and (apparent) magnitudes of the galaxies and we wish to assign each galaxy an initial triplet $(Q_{11},Q_{22},Q_{12})$ of unweighted quadrupole moments. The Cauchy-Schwartz inequality for quadrupole moments implies that $|Q_{12}|$ is bounded by $\sqrt{Q_{11}Q_{22}}$. Thus, the distributions of the moments are not independent of each other and cannot be sampled independently from a marginal distribution as was done in~\cite{2015arXiv151205591I}. Moreover, the shapes and sizes of the galaxies depend on the redshift, magnitude, morphology, etc. Faint galaxies are more likely to be found at higher redshifts and thus may have smaller angular sizes; see for example \citetalias{Massey13}. The polarisation distribution can have a mild dependence on the local environment as well~\citep{Kannawadi15}. 

To learn the joint distribution of the quadrupole moments from real data we use the galaxy population in the COSMOS field as our reference and assign shapes and sizes that are consistent with the observed distribution in the COSMOS sample. Since the unweighted moments are not directly available from the data, we have to rely on parametric models fitted to the galaxies. We use the publicly available catalogue of best-fit \sersic\ model parameters for COSMOS galaxies as our training sample~\citep{Griffith2012a}. 
The catalogue consists of structural parameters such as \sersic\ indices, half-light radii, and polarisation prior to the PSF convolution (this is done in that paper by modelling the PSF at each galaxy position), in addition to magnitudes and photometric redshifts for about 470\,000 galaxies. 

We model the 6-dimensional multivariate distribution of magnitude, redshift, polarisation, half-light radius and \sersic\ index using a mixture of 6D Gaussians. A generative model such as this one has the advantage that we can generate arbitrarily large mock catalogues that are statistically similar to the catalogue we begin with, without having to repeat the values in the original catalogue.  We find that with 100 Gaussian components, we are able to recover the 1-dimensional and 2-dimensional marginal distributions very well. We obtain a mock catalogue, sampled from the Gaussian mixture model, with three times as many entries as the MICE catalogues have. We remove from the mock catalogue any unrealistic values (such as polarisation above $1$ or redshift less than $0$), caused by over-extension of the model into unrealistic regimes. We then find the closest neighbour for each galaxy in the MICE catalogues in magnitude-redshift space using a kd-tree and assign the corresponding polarisations. The orientations of the galaxies are random and uncorrelated with any other parameter, thus any coherent, intrinsic alignment among the galaxies is ignored. The model is hence too simplistic to capture the environmental dependencies on shapes and sizes.

Using the knowledge of circularised half-light radii along with their \sersic\ indices, the $R^2 = Q_{11} + Q_{22}$ values assigned to the galaxies are second radial moments computed analytically for their corresponding \sersic\ model. Additionally, with the knowledge of polarisation and position angle, which are in turn obtained from the best-fit \sersic\ model, we obtain all three unweighted quadrupole moments $(Q_{11},Q_{22},Q_{12})$. 

\subsection{Survey}
\label{sec:survey}
A key feature of our approach is that survey characteristics are readily incorporated. Having assigned the galaxy properties, 
we simulate a $10\times 10$ deg$^2$ survey with a simple scanning strategy. We tile the VIS focal plane following the current design, see Sect. \ref{detector}. 

\begin{figure}
\includegraphics[width=\columnwidth]{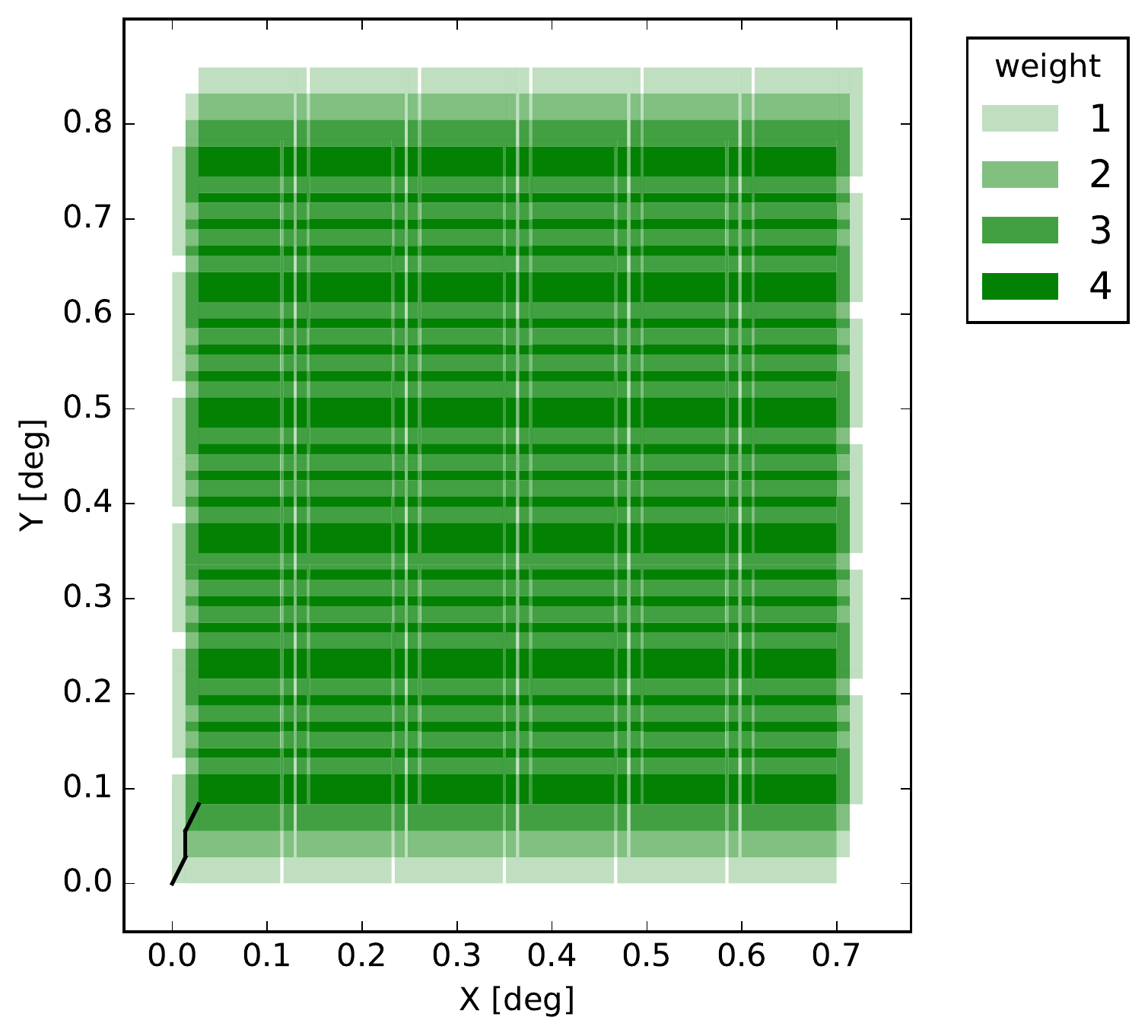}
\caption{Coverage of a single slew by VIS. The default dither pattern in \emph{Euclid} is `S'-shaped (shown as the black lines in the lower left corner) with displacements ($\Delta x$,$\Delta y$)$=$($0$,$0$;\;$50$,$100$;\;$0$,$100$;\;$50$,$100$)$''$. The weights show the number of times an area has been observed. In each field of view there are $6\times 6$ non-square CCDs, with asymmetric spacing between them in the vertical and horizaontal driections, which results in a non-square field of view.}
\label{fig:dither}
\end{figure}

To fill the gaps between its CCDs, Euclid will observe in a sequence of four overlapping exposures that are offset (or `dithered') with respect to each other; a re-pointing between the sets of overlapping exposures, i.e. dither, is called a `slew'. The nominal pattern of offsets for exposures $i=1,\dots,4$ creates an `S'-shaped pattern \citep[see][for more details]{Markovic16}, where the angular shifts with respect to the previous field positions are:
$(\Delta x_1,\Delta y_1)=(50,100)$;
$(\Delta x_2,\Delta y_2)=(0,100)$; 
$(\Delta x_3,\Delta y_3)=(50,100)$
in arcsec. The code uses Mangle \citep{Swanson08} to create the corresponding weight map and tiles this map across the survey patch (the code is flexible enough to incorporate any dither pattern). The weight map for a pointing with four dithers is shown in Fig.~\ref{fig:dither}.

The propagation of the PSF and CTI stages of the pipeline,
and the inverse relations described in Eqs.~(\ref{ref}) and (\ref{pert}), are performed on a per exposure basis. The resulting polarisations are then averaged over all of the exposures that each galaxy receives, subject to the dither pattern (some areas of sky have fewer than four exposures, and this is captured by the dither pattern described here).

We also simulate a simple scanning strategy by ordering the tiling of the survey area in row (right ascension) order followed by column (declination) order, i.e. a rectilinear scanning strategy \citep[see][]{Kitching16}. In future implementations this will be generalised to match the 
full {\it Euclid}  reference survey scanning strategy (Scaramella et al., in prep). 

In this first implementation and presentation of the code we do not include uncertainties in the spatial variation of foreground sources of emission or extinction. However, given the pipeline infrastructure these can be readily included and will be investigated further in future studies. 

\subsection{Instrumental effects}

We limit our analysis to the two main sources of instrumental bias, namely uncertainties in the PSF caused by focus variations and the impact of an imperfect correction for CTI. 

\subsubsection{Point Spread Function (PSF)}

Correcting the observed shapes to account for their convolution by the PSF is an important step in any weak lensing measurement pipeline, and much effort has been spent on the development of algorithms to achieve this. A critical ingredient for the correction is an accurate model of the PSF itself \citep{Hoekstra04}. Current cosmic shear studies take a purely empirical approach where the spatial variation of the PSF is captured by simple interpolation functions that are fitted to the observations. In the case of \emph{Euclid} with its diffraction-limited PSF this is no longer possible: the PSF depends on the SED of the galaxy of interest \citep{Cypriano10,Eriksen18}. Moreover, compared to current work, the residual biases that can be allowed are much smaller given the much smaller statistical uncertainties afforded by the data. Therefore, a physical model of the telescope and its aberrations is being developed (Duncan et al., in prep.). The PSF model parameters are then inferred using measurements of stars in the survey data, supported by additional calibration observations. 

The model parameters, however, will be uncertain because they are determined from observations of a limited number of noisy stars. Constraints may be improved by combining measurements from multiple exposures thanks to the small thermal variations with time. The PSF will, nevertheless, vary with time, and thus can only be known with finite accuracy. Moreover, the model may not capture all sources of aberrations, resulting in systematic differences between the model and the actual PSF. Fitting such an incorrect model to the measurements of stars will result in residual bias patterns \citep[e.g.][]{Hoekstra04}, that may be complicated by undetected galaxies below the detection threshold of the algorithms used for object identification \citep{Mart19}.

The PSF uncertainties in the pipeline are based on the current {\it Euclid} PSF wavefront model and capture one of the main sources of uncertainty, which is the nominal focus position, as detailed in Appendix~A. We note that our results are expected to be somewhat conservative for this particular example, because we ignore the correlations in focus positions between subsequent exposures. On the other hand, a more realistic scenario is expected to introduce coherent patterns on smaller scales caused by errors in the model itself. This will be studied in more detail in future work.

\subsubsection{Detector}
\label{detector}
The VIS focal plane is comprised of $6\times 6$ CCDs that each have dimensions of $(2\times 2048)\times (2\times 2066)$ pixels, where we explicitly indicate that each CCD consists of four separate readout circuits (quadrants). 

Thanks to their high quantum efficiency and near linear response, CCDs are the most practical devices to record astronomical images. They are, however, not perfect and various detector effects can degrade the images. Examples include the brighter-fatter effect \citep[BFE; e.g.][]{Antilogus14,Plazas18}, which affects bright objects such as stars, detection chain non-linearity, offset drifts and photo-response non-uniformity. Here we focus on CTI, caused by radiation damage that accumulates over time in the detectors. The resulting trailing of charge changes the measured shape and has a larger impact on fainter objects, and is, therefore, most damaging for weak lensing studies. 

There is an extensive, ongoing, characterisation programme that focuses on CTI for {\it Euclid}'s detectors, the CCD273 from e2v, \citep[see e.g.][]{2012SPIE.8453E..16G, 2012SPIE.8453E..15H, 2014SPIE.9154E..14P,Niemi2015}. The results from this on-ground characterisation work, together with calibration measurements acquired in flight with the  actual {\it Euclid} detectors, will allow the data processing to mitigate the biases caused by CTI, using correction algorithms such as those described in \cite{10.1093/mnras/stu012}.
There is a fundamental floor to the accuracy of CTI correction, even if the model exactly matches the sold-state effect, owing to read noise in the CCD. The model will also have associated systematic errors and uncertainties that will translate into increased noise and residual biases for the shape measurements, with preferred spatial scales corresponding to those of the quadrants (which are approximately \ang{;3.5;} in right ascension and $4'$ in declination) and the CCDs (which are approximately $7'\times 8'$).  

As there are more electrons from brighter sources, the relative loss of charge due to CTI is lower. As a result, CTI affects fainter and extended sources more \citep[e.g. see Figs.~10 and~11 in][]{Hoekstra11}. In our current implementation, which is detailed in Appendix~B, we ignore these dependencies. Instead we consider a worst case scenario, adopting the
bias for a galaxy with ${\rm SNR}=11$ and FWHM of $\ang{;;0.18}$ and a trap density that is expected to occur at mid survey. These parameters are  based on the results from \cite{Israel15} \citep[with updated parameters as presented in][]{Israel17}, who adopted the same approach.

As discussed in Appendix~B, CTI is expected to increase with time as radiation damage accumulates. To account for this increase, here we assume that trap densities grow linearly with time. This gradual trend is further deteriorated by intermittent steps, which are caused by solar Coronal Mass Ejections (CMEs), which largely increase the flux of charged particles through the detectors over the baseline level. This means the estimate of the trap density parameter has to be updated periodically using images acquired in orbit. To investigate this effect in the model we define `reset on' or `reset off' cases. The two cases affect the estimated trap-densities, $\rho$, and the associated errors in the model. In the first case the relative error in the density of species i, $\delta\rho_i$, is the same throughout the whole patch of the sky under study\footnote{The absolute error in the density of species $i$ is just given by $\Delta\rho_i=\rho_i \times (1+\delta\rho_i$), where $\rho_i$ is the `true' trap density.}, sampled from a normal distribution with zero mean and standard deviation $\sigma_p$. Hence, for each realisation all measurements in the observed patch are affected by the same relative error in trap-density; we refer to this case as `reset off'. 

The second case is `reset on', where we model the potential effect of resetting the CCD after a CME event, a so-called `CME jump' on scales smaller than those of the considered patches. In this case the relative error in trap densities are re-estimated midway through the patch, meaning it has one value in one half of the patch and another in the other half, both drawn from the same distribution as that used in the `reset off' case. And again these biases are updated (sampled from the same normal distribution) in every realisation. This scenario would correspond to a more frequent, but equally accurate, update of the trap-densities than the `reset off' case and the coherence of the biases across the angular scales is decreased by the jumps, or resets, across the patch halves. The point is that the error is never exactly zero. But one will have to re-do the model in the case of a CME jump that will cause a different model uncertainty. 

\subsection{Power spectrum computation}
For each realisation we take a spherical {\tt HEALPix} map of the galaxies to make an estimate of the shear map for both the reference and perturbed catalogues. The unobserved areas are masked, and we apodise this mask with a Gaussian with a standard deviation $\sigma=1.5\pi/2048$ to minimise the effect of the result of leakage due to the boundaries. We then use {\sc anafast} from {\tt HEALPix} to calculate the $E$-mode power spectrum of the masked map.

\subsection{Pipeline setup}
A key feature of our approach is that we create realisations of the systematic effects, for each galaxy and each pointing, which enables us to determine the expected probability distributions for the changes in the cosmological parameter inferences caused by these systematic 
effects. This is done by creating 150 random realisations that are propagated through the Fisher matrix and bias calculations as discussed in Sect. \ref{cosmo}; we choose $150$ since this then means the total area is $150\times 100$ square degrees which is equal to the total \Euclid wide survey. The run where we combine PSF and CTI residuals took 20 hours to compute on a machine with 25 1.8\,GHz CPUs and 6\,GB RAM.
The PSF-only scenario took 14 hours, and the CTI-only run took seven hours on the same architecture. As each realisation can be run in parallel, the calculations can be sped up accordingly on a machine with more processors.

\begin{figure*}
\includegraphics[width=\columnwidth,height=7cm]{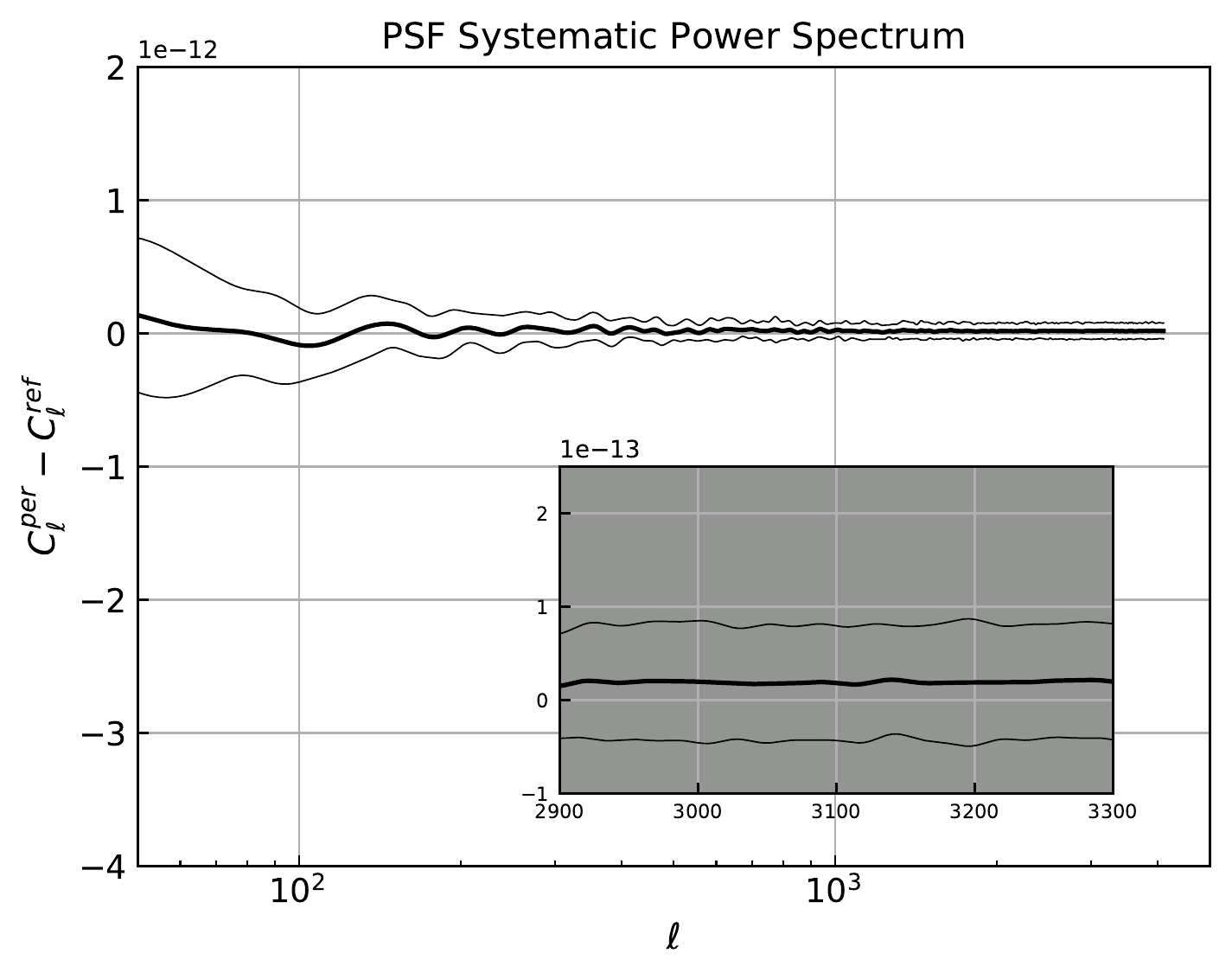}
\includegraphics[width=\columnwidth,height=7cm]{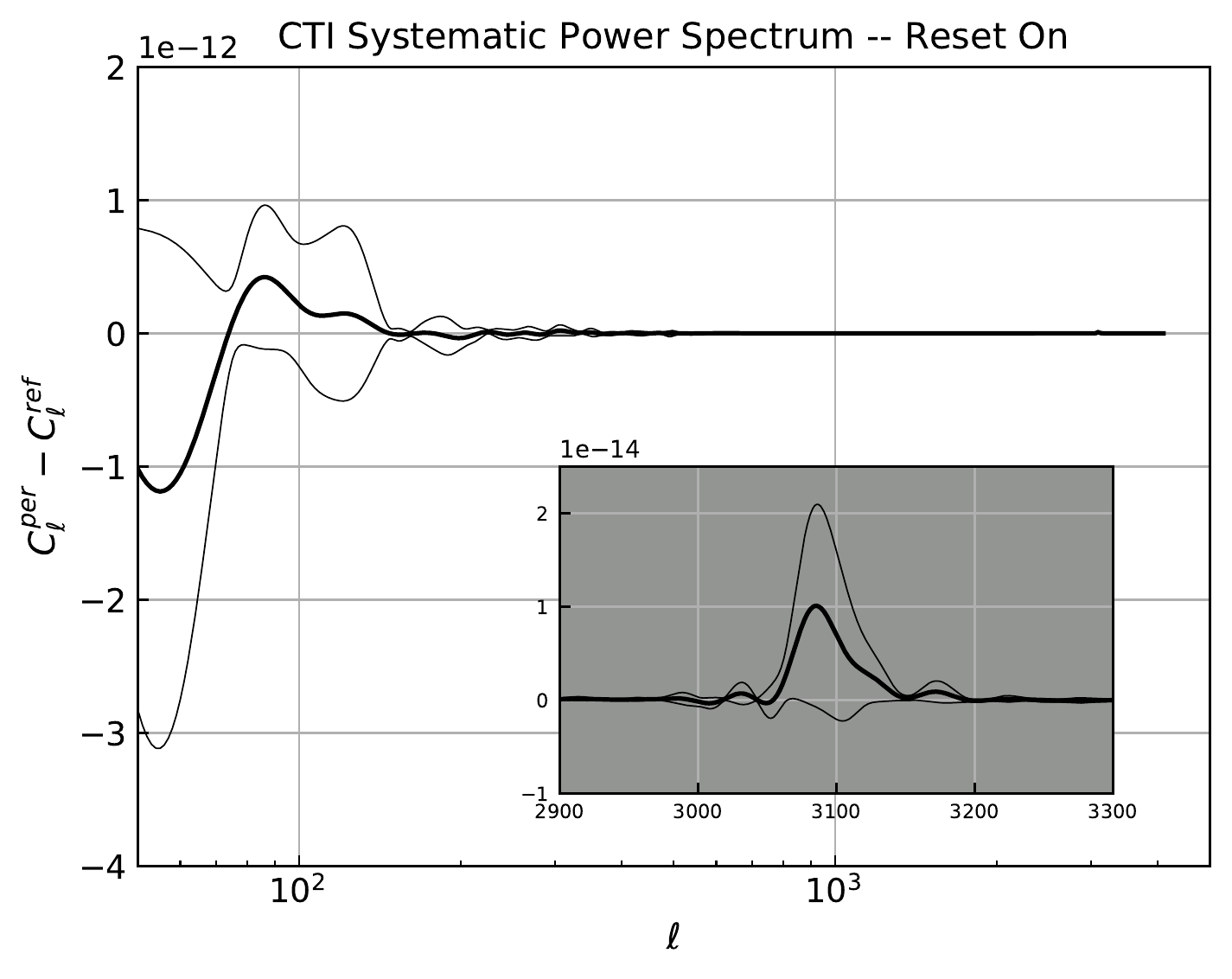}
\includegraphics[width=\columnwidth,height=7cm]{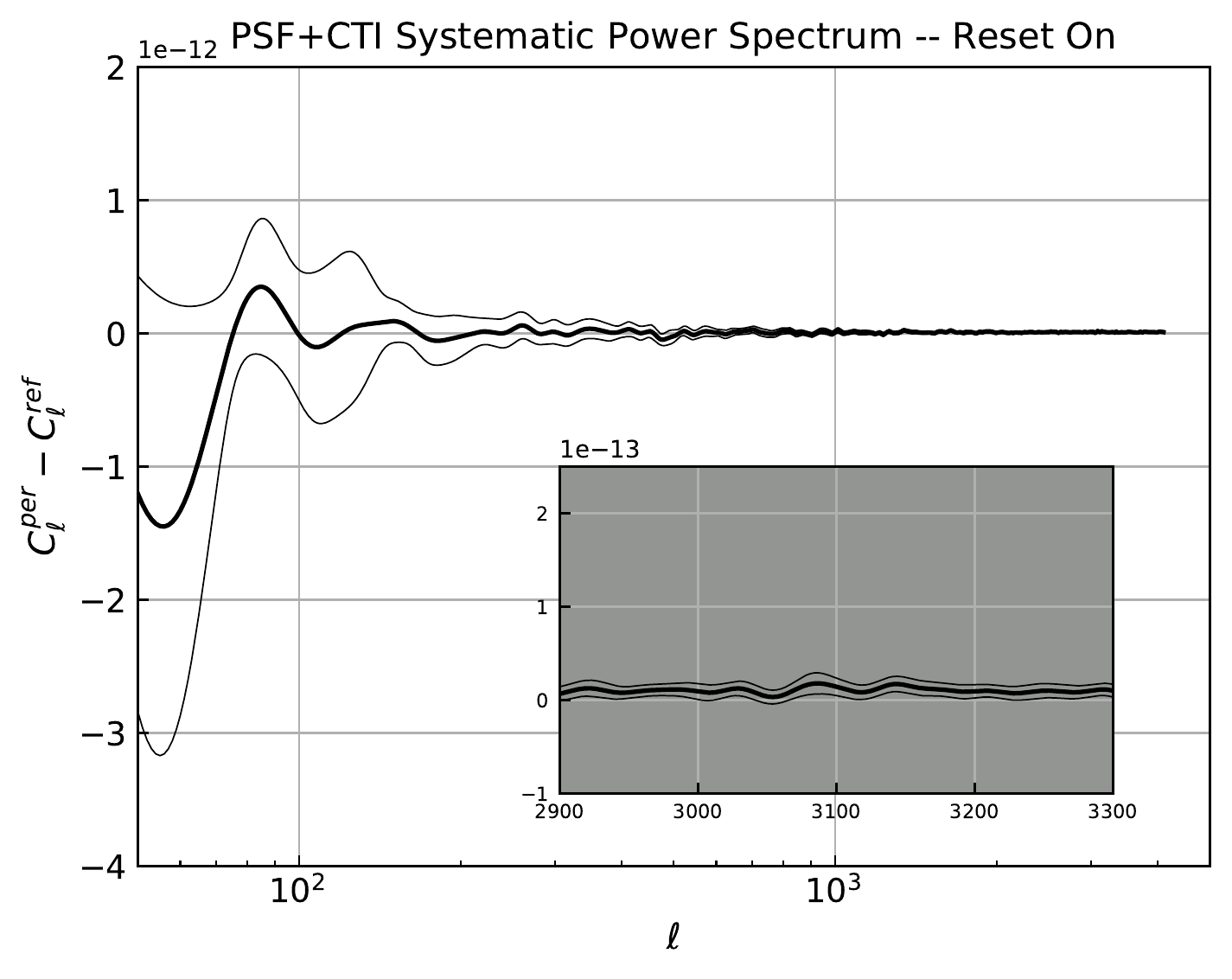}
\includegraphics[width=\columnwidth,height=7cm]{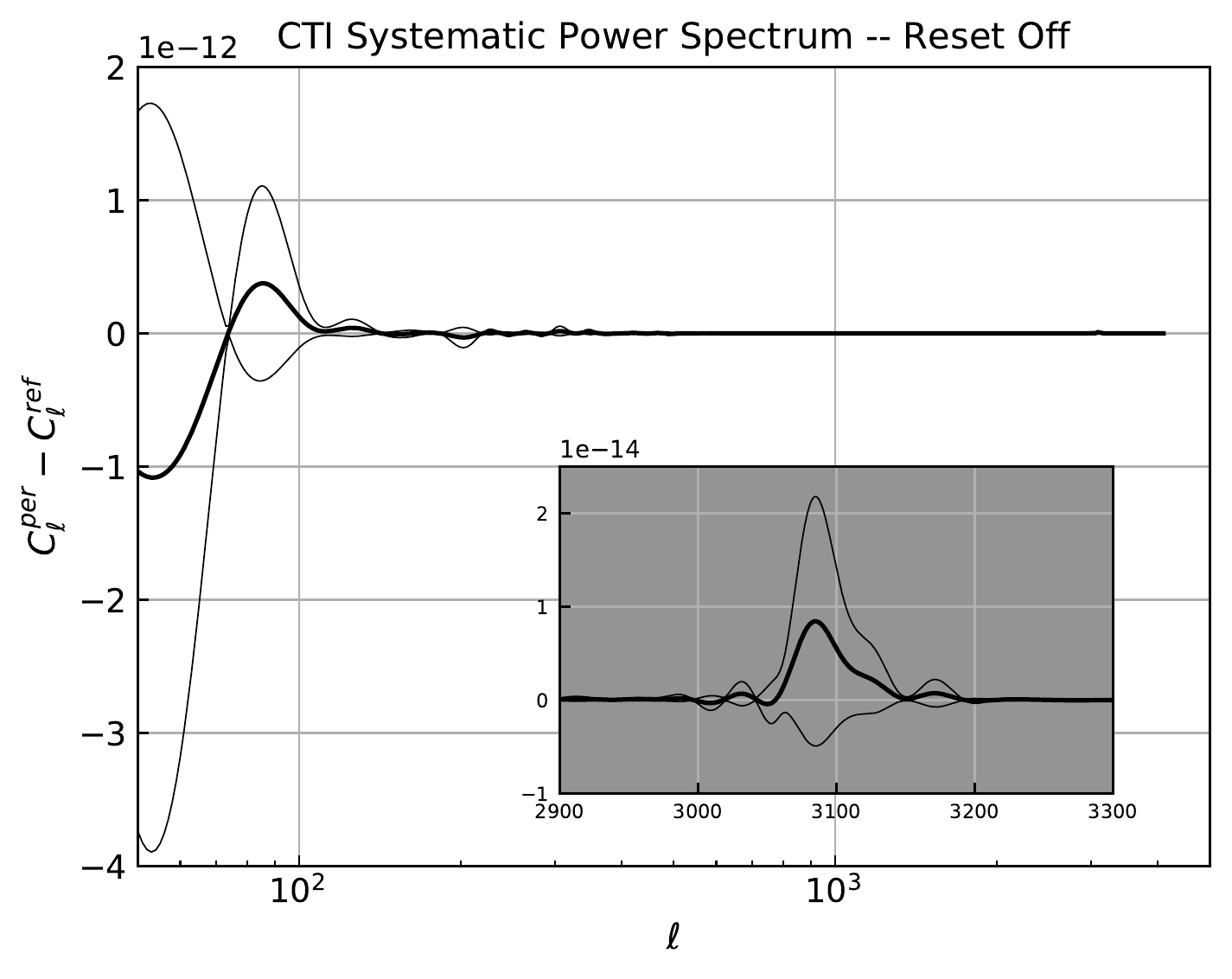}
\caption{Residual power spectra caused by imperfect removal of systematic effects. Thin lines show $68\%$ intervals. The upper left panel shows the residual power spectrum due to PSF, caused by the limited precision with which the nominal focus position can be determined from the stars in the data; it can be seen that residuals have on average been removed. The upper right panel shows the residual power spectrum caused by CTI when the CTI-removal model parameters are updated throughout the survey (`reset on' case, see text for details). There are residuals on the scales corresponding to half the distance between the CCDs, as shown in the insets. The lower right panel shows the results when the CTI-removal model parameters are kept constant during the survey (`reset off'). As can be seen the residuals have a slightly wider distribution compared to the `reset on' case. The lower left panel shows residual systematic effects from uncertainties in the modelling of both PSF and detector effects; as shown in the inset the two effects seem to work in opposite directions where the positive offset present in the PSF-only case has reduced in the combined case. We note that due to the sensitivity of dark energy parameters to relatively large angular scales $\ell\simeq 50-1000$, the deviations on these scales are of more importance.}
\label{fig:sys_cls}
\end{figure*}

\section{Results}
\label{sec:casestudies}

As a demonstration of the usefulness of our approach, 
we assess the impact of two prime sources of bias for the {\it Euclid} cosmic shear analysis: PSF and CTI 
modelling. We compute the expected residual systematic power spectra caused by imperfect removal of systematic effects from realistic uncertainties in the modelling. We then propagate the power spectrum residuals through a Fisher matrix to compute the biases in dark energy parameters. 

\subsection{PSF}
\label{sec:psf}

The upper left panel of Fig.~\ref{fig:sys_cls} shows the residual systematic power spectrum caused by uncertainties in the PSF model caused by focus variations. The thick line indicates the mean of the 150 realisations, whereas the thin lines delineate the 68\% interval. As discussed in Appendix~A, we consider only the uncertainty in the PSF model given the assumed nominal focus position, which is the dominant contribution and introduces residuals in the power spectrum on large scales. Other imperfections in the optical system will typically introduce residuals on smaller scales.

\begin{figure}
\includegraphics[width=\columnwidth]{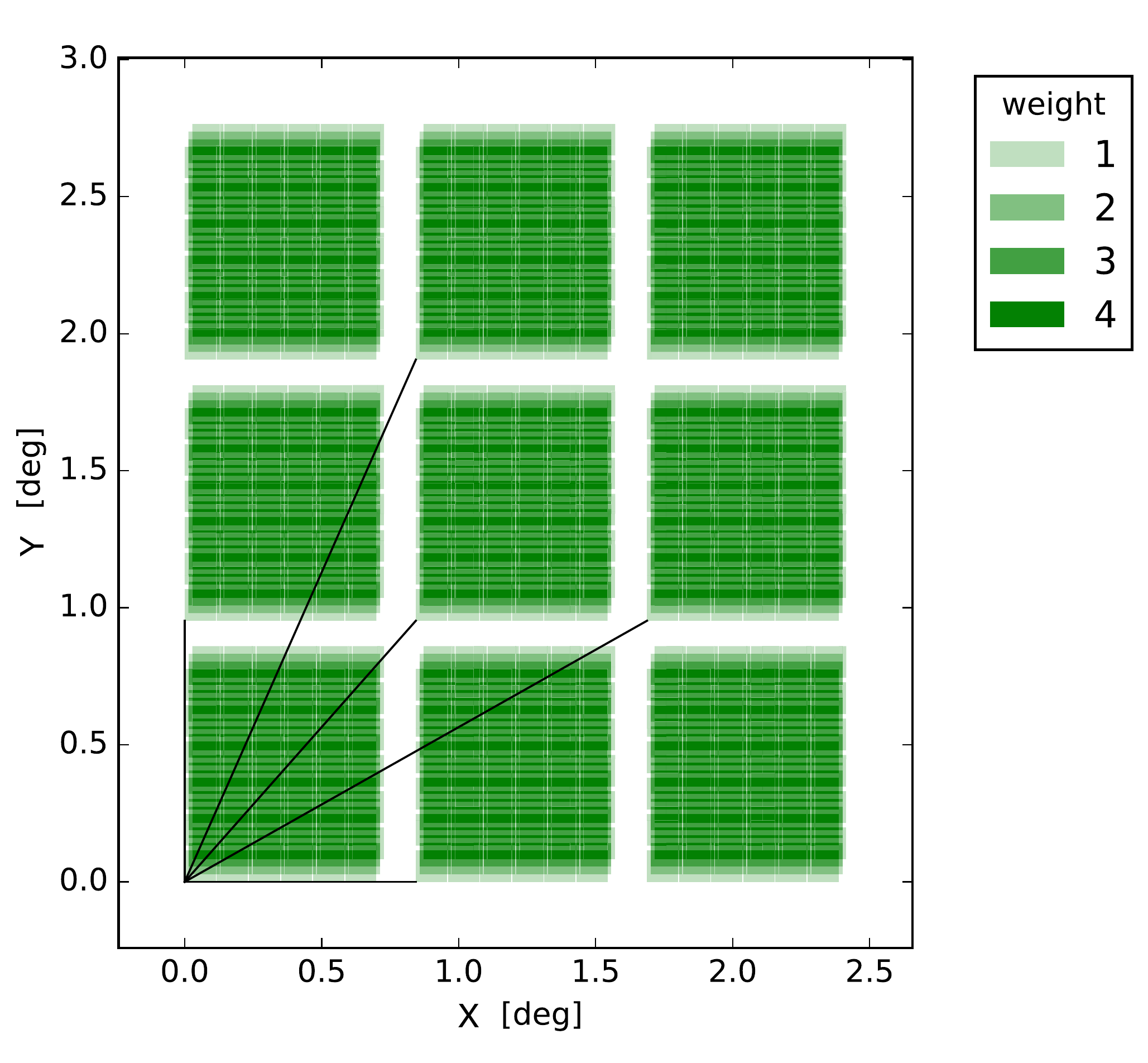}
\caption{Part of the observed area with $3$ slews in each direction and four dithers for each slew. The slews are 
plotted at $1.2\times$ their nominal value for presentation purposes, causing apparent gaps, which are not present in the actual simulated survey. The lines show some of the correlated scales relating to the same point in each field of view. We also note that there are correlations at $2\times,3\times,n\times$ of these harmonic scales. It should be noted that relevant scales are determined by the {\it distances} between the fields-of-view, not the size of the field-of-view itself.}
\label{fig:FoV}
\end{figure}

To understand the relevant scales in the PSF case, it is helpful to look at Fig.~\ref{fig:FoV}, where some of the relative correlated scales are indicated. A point in one field-of-view is correlated with the same point in all the other fields-of-view -- i.e. the angular {\it distances} between the fields-of-view are also relevant here, not only the scales of field-of-view itself. Also the field-of-view is not square, and hence the distances to the same point in the fields-of-view are not the same in both directions. In our $10\times 10\,{\rm deg}^2$ area, this gives us a {\it range} of correlated scales: $13 \le \ell \le 300$. The minimum distance between adjacent fields-of-views corresponds to $\ell=300$, and the diagonal in our square survey area (the maximum angular separation)corresponds to $\ell=13$. Incidentally this is also the range where cosmic variance dominates. 

The average residual power spectrum in the top left panel of Fig.~\ref{fig:sys_cls} is close to zero and does not show sharp features, but the residual PSF biases contribute over a {\it range} of scales. This is because the averaging over the four dithers for each slew reduces the average induced biases in the polarisations, which in turn reduces the correlations between slews; and the polarisations in the perturbed line for each field-of-view (i.e. each dither and each slew) are drawn from a distribution, so that the average impact is typically less extreme.

\subsection{CTI}
\label{sec:cti}
The thick line in the top right panel in Fig.~\ref{fig:sys_cls} shows the average residual power spectrum when we consider the imperfect correction for time-dependent CTI for the `reset off' case (see Sect. \ref{detector}). The amplitude of the residuals are slightly larger than that of the PSF case. Compared to the PSF case, there are additional angular scales on which correlations can occur, namely the distances between the CCDs in the detector. The inset shows a zoom in around $\ell\simeq 3080$, which corresponds to half the distance between CCDs. This is because in our setting, CTI systematic effects are induced only in the serial readout direction (see Appendix~B), inducing biased polarisation estimates at half the CCD scale (quadrant scale).

In the second case, `reset on' (see Sect. \ref{detector}), the results presented in the bottom right panel of Fig.~\ref{fig:sys_cls} show that this procedure does not improve the residuals around $\ell\simeq 3080$. It does, however, reduce the variance on the largest scales, even though the average residual power spectrum is largely unchanged, except for increased variation for $\ell$ in the range $150-300$.

\begin{figure*}
\includegraphics[width=0.999\columnwidth]{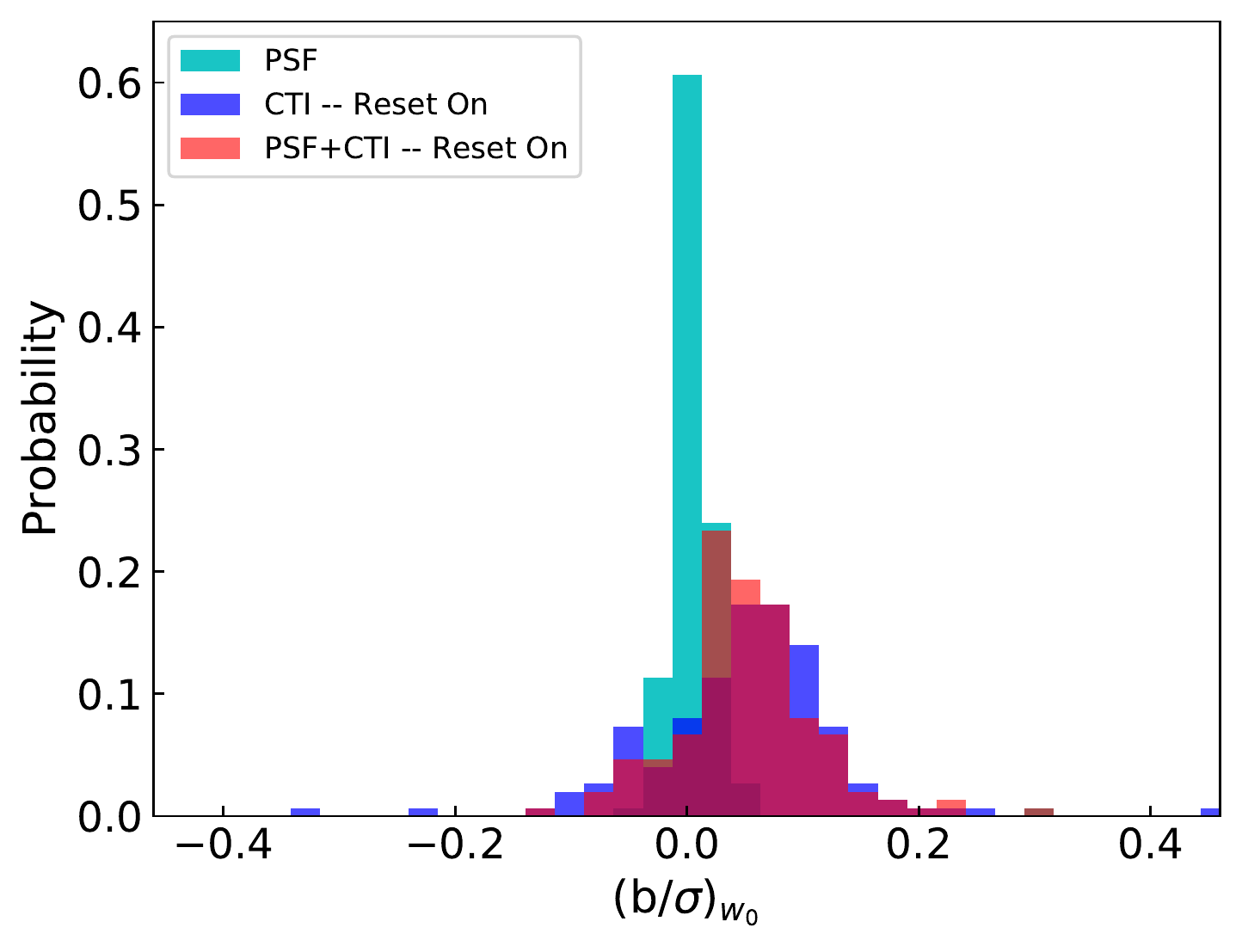}
\includegraphics[width=.999\columnwidth]{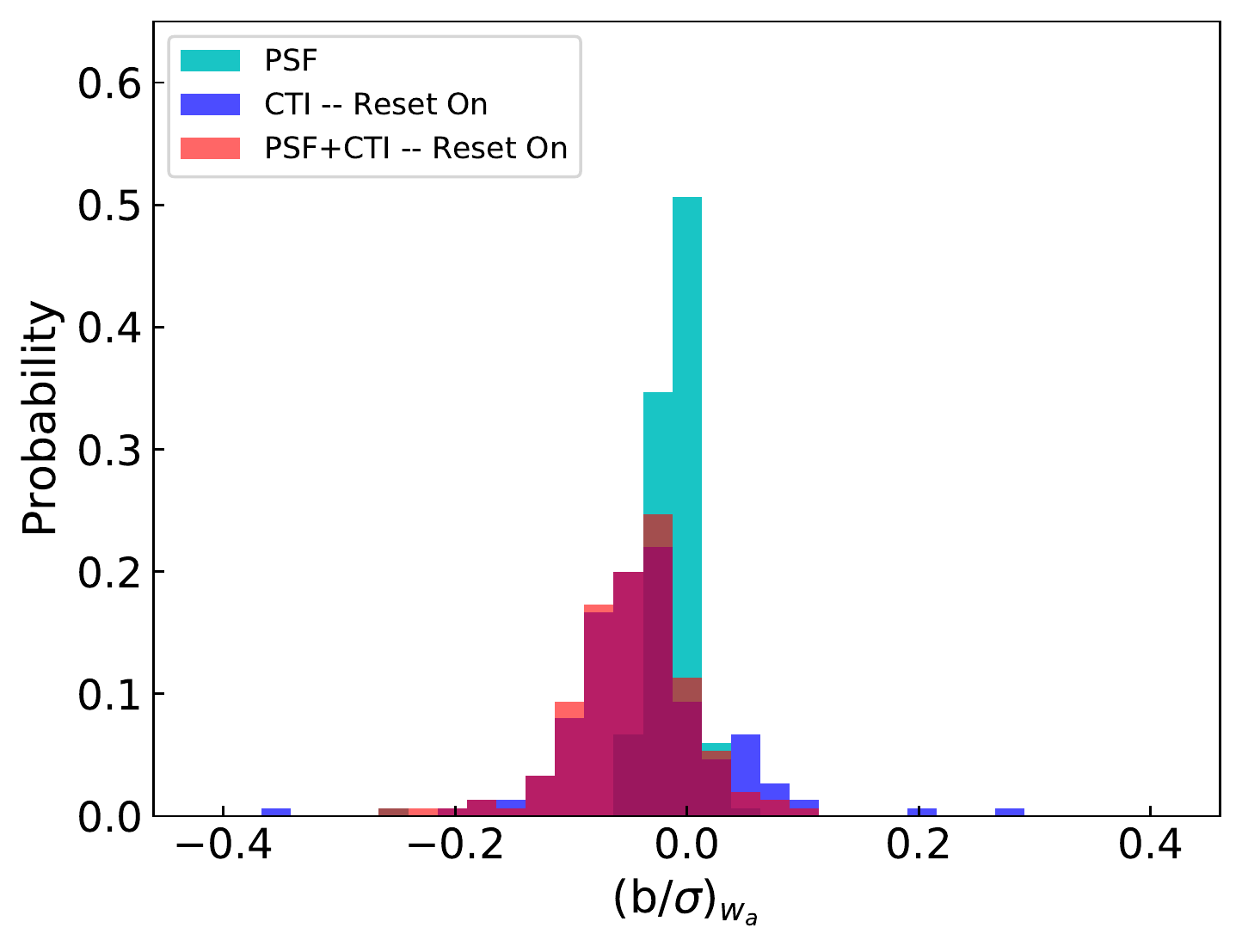}
\caption{{\it Left panel:} ratio of the bias in $w_0$ and the $1\sigma$ uncertainty in this parameter for PSF-only (cyan), CTI-only with resetting on (blue) and both PSF and CTI with resetting on (red) scenarios. {\it Right panel:} ratio of the bias in $w_a$ and the $1\sigma$ uncertainty in this parameter. Although the distributions are wide in some scenarios, we find that they are well within limits set in \citetalias{Cropper13} -- also see Table~\ref{scenes}.}
\label{fig:stats}
\end{figure*}

\subsection{PSF and CTI}
\label{sec:psf_cti}

Rather than considering individual sources of bias separately, we can simultaneously propagate different types of systematic effects and capture their correlated effects. This is demonstrated in the bottom left panel of Fig.~\ref{fig:sys_cls}, which shows the residual systematic power spectrum resulting from both CTI (reset on) and PSF systematic uncertainties\footnote{Here we ignore the impact that CTI can have on the PSF measurement. However this is expected to be a small effect; see lines 2\&4 of Table 1 in \cite{Israel15}.}. Both features of CTI and PSF systematic effects can be seen in the residual power spectrum. The inset shows the residual power spectrum in the range corresponding to the CCD scales, where CTI contributes most. The residuals on these scales are now dominated by both the CTI and PSF systematic effects.

\subsection{Impact on cosmology}
\label{sec:cosmology}
For each residual power spectrum we compute the change in the expected maximum likelihood locations for the parameters $w_0$ and $w_a$. The tolerable range for biases on dark energy parameters is generically $|(b/\sigma)_{w_0}|\le 0.25$ (where $b$ is the bias, and $\sigma$ is the 1-$\sigma$ marginalised uncertainty) as derived in M13 and \cite{2018PhRvD..98d3532T}, which ensure that the biased likelihood has a greater than 90\% overlap integral with the unbiased likelihood.

The results are presented in Fig.~\ref{fig:stats} and reported in Table~\ref{scenes}. We show results for the PSF-only case (cyan), the CTI-only case with resetting on (blue), and the combined case (red). The panels respectively show the biases in $w_0$ and $w_a$ relative to the statistical uncertainty. In Table~\ref{scenes} we list the mean and its uncertainty for the quantities, as well as the standard deviation of the distributions themselves. We also quote the 90\% confidence limits of the bias distributions. 

We find that the PSF residuals have a minimal impact, which is expected as the amplitudes of the residual power spectra were small.
The induced biases $b$, relative to the uncertainty $\sigma$ on the dark energy parameters are expected to be $(b/\sigma)_{w_0} = [-0.024, 0.033]$ and $(b/\sigma)_{w_a} = [-0.042, 0.015]$ at 90\% confidence interval. These are well within the tolerable range.

For the case where the CTI model parameters are kept fixed during the simulated observations of a 100 deg$^2$ patch (`reset off'), the impact on the induced biases are $(b/\sigma)_{w_0}= [-0.328, 0.077]$ and $(b/\sigma)_{w_a}= [-0.054, 0.281]$, which are just outside the tolerable range. However for the case where we resample the CTI model parameters (`reset on'), the results are  improved with $(b/\sigma)_{w_0}=[-0.078, 0.152]$ and $(b/\sigma)_{w_a}= [-0.121, 0.067]$. The effects seen here are very similar to effects seen using the simplified models of CTI in \cite{Kitching16}.  

Perhaps most interesting are the results for the case where we include both CTI and PSF residuals, since this joint case  was not captured in the \citetalias{Cropper13} flowdown. We find that the biases are expected to be $(b/\sigma)_{w_0}=[-0.046, 0.144]$ and $(b/\sigma)_{w_a}= [-0.124, 0.032]$; again within the tolerable range.

\begin{table*}
\begin{tabular}{|l||c|c||c|c|c|}
\hline
&\multicolumn{2}{c|}{Statistics}& \multicolumn{2}{c|}{90\% Confidence Interval} \\
\hline
Effect(s) & $(b/\sigma)_{w_0}$ & $(b/\sigma)_{w_a}$ & $(b/\sigma)_{w_0}$ & $(b/\sigma)_{w_a}$\\
\hline
\hline
PSF & $0.006\pm 0.002(0.029)$ & $-0.018\pm 0.005(0.064)$ & $(-0.024, 0.033)$ & $(-0.042, 0.015)$ \\
\hline
CTI (Reset Off) & $-0.045\pm 0.030(0.370)$ & $0.045\pm 0.027(0.330)$ & $(-0.328, 0.077)$ & $(-0.054, 0.281)$\\
CTI (Reset On) & $-0.049\pm 0.007(0.083)$ & $-0.038\pm 0.006(0.068)$ & $(-0.078, 0.152)$ & $(-0.121, 0.067)$\\
\hline
PSF \& CTI (Reset On) & $0.056\pm 0.006(0.078)$ & $-0.050\pm 0.005(0.066)$ & $(-0.046, 0.144)$ & $(-0.124, 0.032)$\\
\hline
\end{tabular}
\caption{Summary of bias changes for the different case studies. 
The column labelled `Statistics' shows the mean and 68\% error on the mean for our $150$ realisations. The numbers in brackets are the standard deviation of the distributions. The column labelled `90\% Confidence Interval' shows the 90\% confidence regions in our distributions.}
\label{scenes}
\end{table*}

\subsection{Discussion} 
It is useful to compare our findings to the requirements derived in \citetalias{Cropper13}. In the latter study, requirements on systematic effects were set through a formalism that \emph{flowed down} (i.e. subdivided requirements in progressively finer details via a series of inter-related subsystems) changes in the power spectrum parameterised by 
\bee
\label{eq:oldbias}
\delta C(\ell) ={\cal M}\,C^{R}(\ell)+{\cal A}.
\eee
Requirements on ${\cal M}$ and ${\cal A}$ were determined for various effects such as PSF and CTI. To compare to this formalism one could naively fit the residual power spectra that we find using such a linear model. However, this would neglect the correct formulation of how to propagate biases into cosmic shear power spectra (Kitching et al., in prep).
 
Therefore to asses the difference between the \citetalias{Cropper13} approach and our approach we need to \emph{flow up} the requirements on the uncertainties set in \citetalias{Cropper13} (referred to as $\sigma$ in that paper) for individual effects, and compare the outcome of the two approaches at the level of biases in cosmological parameters rather than comparing ${\cal M}$ and ${\cal A}$ values. We do this by determining the multiplicative and additive biases, ${\cal M}$ and ${\cal A}$, associated with each systematic effect in \citetalias{Cropper13}, constructing Eq.~(\ref{eq:oldbias}) for these values, and then adding this to Eq.~(\ref{eq:bias}); a process we refer to a `flow up'.

Whilst uncertainties are included in this flow-down approach, these are taken to be constant across the survey (both spatially and temporally). They are also assumed to be independent of each other. Our approach does not suffer from these limitations. By modelling biases simultaneously, they also have a chance of acting at different scales, or even cancelling each other out. Hence any comparison with prior work should not be interpreted as there being margin in previously derived requirements. Nevertheless, such a comparison is useful to show how different the approaches are, and if previous requirements were exceeded this would be of concern.

Assuming PSF modelling errors in the shear power spectrum at the maximum values permitted by the \citetalias{Cropper13} requirements of ${\cal A}=5\times 10^{-8}$ and ${\cal M}=4.8\times 10^{-4}$, we find biases on cosmological parameters $(b/\sigma)_{w_0}=0.25$ and $(b/\sigma)_{w_a}=0.31$. Assuming CTI correction biases at the maximum values permitted by \citetalias{Cropper13} of ${\cal A}=1.21\times 10^{-8}$ and ${\cal M}=0$ (CTI contributions to multiplicative bias are subdominant) yields $({b}/\sigma)_{w_0}=0.14$ and $({b}/\sigma)_{w_a}=-0.2$. In contrast, our flow-up analysis predicts biases on cosmological parameters that are lower by a factor between $2$ and $5$. None exceed previously derived requirements, and all are within acceptable tolerances to meet top-level scientific goals.

Finally we emphasise several assumptions in this analysis that should be relaxed in future, that may mean the results are either optimistic or pessimistic: 
\begin{itemize}
    \item We do not model intrinsic alignments, the environmental dependence on galaxies' intrinsic size and shapes. 
    \item The smooth increase in CTI over adjacent pointings may be considered optimistic, if CTI has sudden jumps in reality. Furthermore the choice of 45\%-55\% end-of-mission radiation dose is average. In a tomographic analysis CTI residuals may also mimic redshift-dependence of cosmic shear, which may mean the results here are optimistic.
    \item The uncorrelated PSF residuals between consecutive exposures are may be conservative or optimistic, depending on the final state of the telescope at launch. 
\end{itemize}

\section{Conclusions}

We have presented an `end-to-end' approach that propagates sources of bias in a cosmic shear survey at a catalogue level. This allowed the capture of spatial variations, temporal changes, dependencies on galaxy properties and correlations between different sources of systematic and stochastic effects in the pipeline. We use our methodology to revisit the performance of a {\it Euclid}-like weak lensing survey. We limit the analysis to quantify the impact of imperfect modelling of the PSF and CTI, as these are two major sources of bias. Other effects can be readily included, which will be done in future work.

The PSF systematic effects are introduced through the expected uncertainty in fitting the PSF model to noisy data given the assumed nominal focus of the telescope. Additional imperfections will introduce residuals on smaller scales, but these should not affect our main conclusions because the dark energy measurements are most sensitive to variation on large scales. We also consider a time-dependent CTI, where the CTI increases with the survey time due to accumulation of radiation damage on the detectors. We considered a conservative scenario, because the parameters we adopted apply to the faintest galaxies in the analysis, whereas the biases will be smaller for brighter objects. We also do not include intrinsic alignment effects, or source blending effects both of which will be included in future studies. 

These effects were propagated through to residual cosmic shear power spectra and cosmological parameters to estimate the expected biases in the parameters $w_0$ and $w_a$. Compared to requirements based on a more restricted flow-down approach by \citetalias{Cropper13} we find that the biases on the dark energy parameters from our more realistic performance estimates are well within the requirements. Even for the combined scenario of CTI and PSF we find the biases on dark energy parameters are well within the required tolerances.  

This paper presents the first step towards a more comprehensive study of the performance of a {\it Euclid} cosmic shear survey. The same approach, however, can also be readily applied to other cosmic shear surveys. In future work we will introduce more complexity in the PSF and detector systematic effects, so that the resulting redshift dependencies of these effects can be assessed. As alluded to earlier, CTI is dependent on flux and morphology, which implies it will change with redshift. Other systematic effects, such as shape measurement uncertainties, will also be implemented in the pipeline. These improvements will enable us to examine the impact of systematic effects on an increasingly realistic tomographic analysis. 

\begin{acknowledgements}
PP is supported by an STFC consolidated grant. CW is supported by an STFC urgency grant. TDK is supported by a Royal Society University Research Fellowship. HH acknowledges support from Vici grant 639.043.512 and an NWO-G grant financed by the Netherlands Organization for Scientific Research. LM and CD are supported by UK Space Agency grant ST/N001796/1. VFC is funded by Italian Space Agency (ASI) through contract Euclid\,-\,IC (I/031/10/0) and acknowledges financial contribution from the agreement ASI/INAF/I/023/12/0. We would like to thank J\'{e}rome Amiaux, Koryo Okumura, Samuel Ronayette for running the ZEMAX simulations. AP is a UK Research and Innovation Future Leaders Fellow, grant MR/S016066/1, and also acknowledges support from the UK Science \& Technology Facilities Council through grant ST/S000437/1. MK and FL acknowledge financial support from the Swiss National Science Foundation. SI acknowledges financial support from the European Research Council under the European Union's Seventh Framework Programme (FP7/2007--2013)/ERC Grant Agreement No. 617656 ``Theories and Models of the Dark Sector: Dark Matter, Dark Energy and Gravity.

\AckEC

\end{acknowledgements}

\bibliographystyle{aa}
\bibliography{biblio.bib} 

%


\begin{appendix} 
\section{Details of PSF modelling}
In this section, we describe the propagation of uncertainties which result from inaccuracies in the PSF model, 
using the broadband parametric phase retrieval method from Duncan et al. (in prep). 
In this method, the PSF variation is modelled in the wavefront domain. The corresponding real-space optical PSF is obtained 
as the modulus squared of the Fourier transform of the wavefront at the exit pupil, including the effect of telescope distortion,
integrated over the bandpass. In the present application, a simple Gaussian model for the distribution of telescope guiding accuracy is used, linear detector effects (including pixelisation and linear charge diffusion) are included, and it is assumed that the detectors lie precisely in the focal plane. More realistic guiding and detector effects will be included in the future.


The wavefront at the exit pupil of the telescope describes the coherent perturbations in the optical path differences 
of infalling photons, caused by the design and alignment of the telescope optical elements. The wavefront can be split into two parts: an amplitude component, which describes where the light is vignetted by structures in the telescope, 
and a phase component, which describes the variation of the optical path differences. Both change with position 
in the focal plane. To capture the amplitude variation, we use a geometric model that describes 
the projection of intervening structures in the telescope (i.e. the secondary mirror M2 and its struts) at the focal plane. To model the phase variation, we use a suite of simulated wavefronts obtained with
the optical design program {\tt ZEMAX}\footnote{\url{https://www.zemax.com}}, 
configured to the specifications of {\it Euclid}. Each phase map was fitted by a sum of Zernike polynomials, and the variation 
of the corresponding Zernike coefficients with focal plane position was captured by a set of polynomials. Several 
optical elements in the telescope design were displaced or deformed by turn, and the corresponding effects on 
the phase maps were captured by so-called telescope modes. As a result, the wavefront can be predicted for any telescope 
set-up, with a realistic focal plane variation. Given a model wavefront, the real-space PSF is then computed for a range of 
densely sampled wavelengths. The final PSF is obtained by integrating over the spectral telescope response, weighted by the spectral energy distribution (SED) of the source, with additional convolution effects of guiding and CCD pixel response included.
In this application of the model, detector offset and high frequency contributions such as those arising from surface errors are not included.
\begin{figure}
\centering
\includegraphics[width=\columnwidth]{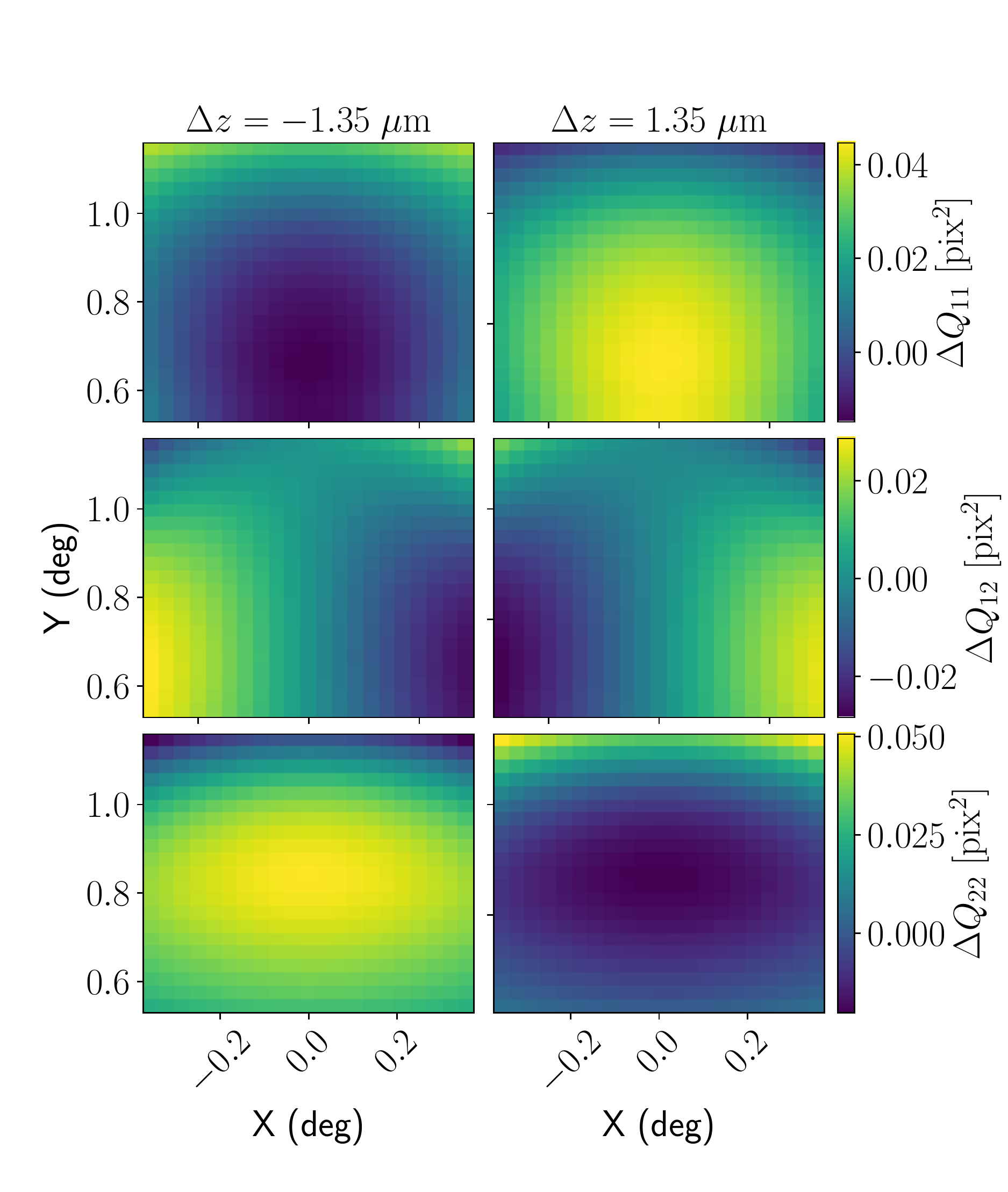}
\caption{The change in the quadrupole moments in units of Euclid pixels squared, for focus position shifts 
of $\Delta z=-1.35\,\mu$m  (left columns) and $\Delta z=1.35\,\mu$m (right columns), for $Q_{11}$ (top panels), $Q_{12}$ 
(middle panels) and $Q_{22}$ (bottom panels) as a function of field-of-view position (in degrees).}
\label{fig:psf_q}
\end{figure}
\begin{figure}
\centering
\includegraphics[width=.9\columnwidth]{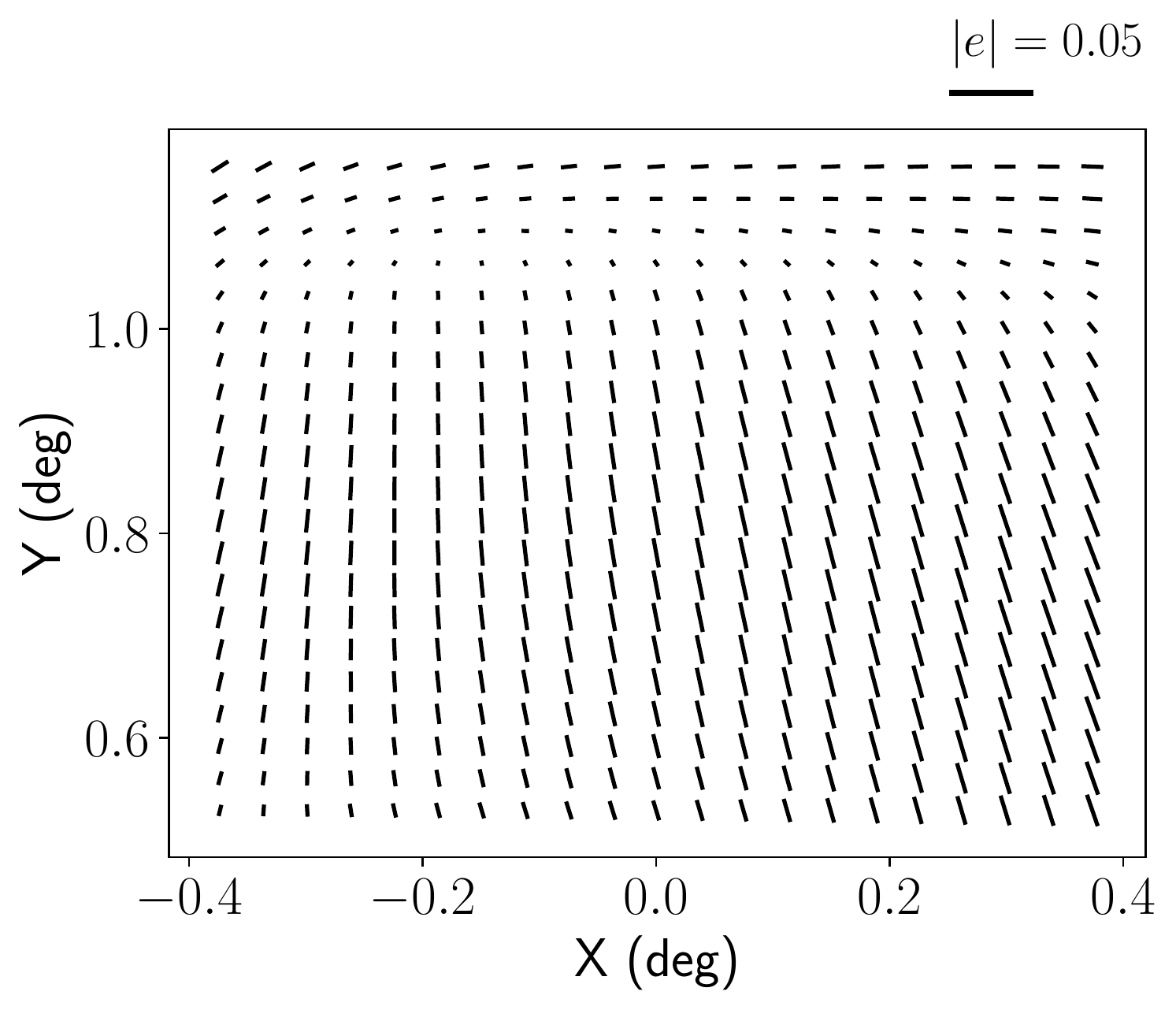}
\caption{Stick plot describing the variation in polarisation across the field-of-view (in degrees) for the nominal focus ($z=0$). All values shown here are taken from the training data, not the fit.}
\label{fig:psf_stick}
\end{figure}

As the Euclid VIS PSF model is jointly fitted to stars in the entire field of view, PSF errors are correlated across that field.  In order to capture this, we investigate the effect of varying one of the principal model parameters, the wavefront error associated with defocus of the telescope.  Higher-order wavefront errors are also expected to contribute to the PSF uncertainty, but the PSF variations of this mode should be a realistic representation of the actual correlated PSF errors (in the absence of possible effects at cryogenic temperatures that could cause deformation in the y-axis displacement of M2 rather than the z-axis). We model the effect of a shift in the focus position resulting in an optical defocus for a given source SED. We choose the source SED to be the template spiral Sbc galaxy of \cite{Coleman80}, with redshift of $1$.

We assume a nominal offset focus position that is drawn at random from a normal distribution whose variance matches the expected $\sigma_z \simeq 0.5\,\mu$m uncertainty in this model parameter, that can be obtained from fitting the telescope model to the stars that appear in each survey field. Even consecutive exposures are assumed to have independent nominal focus values. 

A realisation of errors in the quadrupole moments of the PSF was then obtained in a two-step process. In the first step, the focus position was shifted to give a minimum in PSF size at the centre of the field-of-view (FoV), which was taken as the nominal in-focus position. The mirror was then perturbed in both
directions (positive and negative $z$ offsets) until the second-moment measure of the model PSF's size, $R^2$, varied from the nominal value $R^2_\mathrm{nom}$ by a tolerance $\Delta R^2\equiv |R^2-R^2_\mathrm{nom}| < 10^{-3} R^2$. This value is both the requirement on the knowledge of the mean PSF size across the survey, set by \citetalias{Cropper13}, and also is approximately the measurement uncertainty that we expect to obtain from measurements on individual survey exposures.

In a second step, the FoV-dependent behaviour was determined by fitting a 6\textsuperscript{th}
order polynomial\footnote{This was found to best fit the spatial variation in terms of a least squares minimisation when varying the polynomial order, although the polynomial model itself was assummed.} to $\Delta Q_{ij}(x,y,\Delta z) = Q_{ij}(x,y,z) - Q_{ij}(x,y,z_{\rm nom})$ across $x$ and $y$, where $x$ and $y$ describe the FoV position, $z$ the focus mirror position, $z_{\rm nom}$ the nominal focus position and $\Delta z = z-z_{\rm nom}$. The variation with $z$ was modelled by training the coefficients of the FoV position fit across $\Delta z = 0 \pm 1.35 \,\mu$m, using a quadratic form. Finally, it was verified that the fit recovered the expected perturbed quadrupole moments to $\lesssim 5\%$ accuracy across the field of view, except where $\Delta Q$ was close to zero so that
relative differences -- defined as $(\Delta Q^{\rm fit}_{ij}-\Delta Q_{ij})/\Delta Q_{ij}$ -- became large owing to numerical inaccuracy.
This verification was conducted both on the training data, as well as a coarse grid of field-of-view values and M2 shifts within the $\sigma_z = 0.5\,\mu$m range, which was not used to train the fit. The residual quadrupole moment variations that these changes induce is
shown in Fig.~\ref{fig:psf_q}. Fig.~\ref{fig:psf_stick} shows the variation in polarisation across the field-of-view at the nominal focus position ($z=z_{\rm nom}$).

\section{Details of CTI modelling}
\label{app: CTI}
Charge Transfer Inefficiency (CTI) is caused by the capture and delayed release of photoelectrons by `traps', i.e.\ localised, unintended quantum levels in a detector's silicon lattice. These defects are created when high energy particles displace silicon atoms and, above the protection of the Earth's atmosphere, will accumulate throughout the mission. The timescales for capture and release depend upon the type of damage, and the proximity to any lattice impurities. During readout, if electrons are captured from a charge packet that is moving through the lattice, and released after a sufficient delay, they will become part of a later charge packet (or pixel in the resulting image). This creates faint luminous `trails' behind the images of galaxies and stars, which bias measurements of their polarisation and size. Objects farther from the readout nodes gain brighter trails, because their electrons must travel farther, and are subject to more traps.

We adopt the \cite{Israel15} model of CTI in the {\em Euclid} VIS serial readout direction\footnote{No accurate model yet exists of CTI in the {\em Euclid} VIS parallel readout direction, due to difficulties with engineering model CCDs. We therefore ignore it here. However, parallel CTI has been measured sufficiently accurately to determine that it is subdominant to serial CTI \citep{Endicott}.}. This treats charge capture as instantaneous, and charge release as a stochastic process governed by exponential decay \citep{10.1093/mnras/stu012}. {\it Euclid} CCDs will contain 3 trap species in the serial register. Each species $i$ has a different characteristic release time $\tau_i$, and a time-evolving surface density (abundance) $\rho_i$. 
Table~\ref{Table:M} shows the trap properties expected after the radiation dose accumulated by the end of the mission (for 90\% of realisations of Solar weather). 

\begin{table}
\caption{Baseline trap model used in this work, for an end-of-mission radiation dose \citet{Israel15}, with densities increased by a factor 4.155, following erratum \citet{Israel17}. }
\center
\begin{tabular}{|l|l|l|l|}
\hline
Baseline Model & $i=1$ & $i=2$ & $i=3$ \\
\hline\hline
Trap density $\rho_i$ [pix$^{-1}$] & 0.083 & 0.125 & 3.95\\
\hline
Release Time $\tau_i$ [pix] & 0.8 & 3.5 & 20\\
\hline
\end{tabular}
\label{Table:M}
\end{table}

To account for the accumulation of radiation damage over time, we assume that trap densities grow linearly over the patch of the sky observed, from 45\% of the values in Table~\ref{Table:M} at one end to 55\% at the other. This is quite conservative for large angular scales as it introduces an increase of approximately 10\% over a $\sim$10 day cycle (i.e.\ the typical amount of time it would take to observe our $100\,{\rm deg}^2$ patch), which is much larger than the milder increase over the actual mission. On the other hand, we have assumed consecutive observations, and hence consecutive increases in CTI at each exposure in this patch. In reality, a $100\,{\rm deg}^2$ patch of the observed area will not have this smooth increase as the exposures that cover it will not, in general, be consecutive. Therefore, this is also a somewhat optimistic approach in that sense. 

{\it Euclid}'s baseline strategy for CTI mitigation is a pixel-by-pixel movement of flux from trails, back to the pixels it came from. This `back-clocking' approach is limited by read-out noise (RON), model parameter uncertainties, and model inaccuracy. Because RON is added at the amplifier, it is not trailed during readout -- but it is spuriously corrected by pixel-level methods as if it {\em had} been trailed \citep[see Sect.~5.3 of][]{Israel15}. 

In the reference case (i.e.\ assuming a perfect CTI model), the model and model parameters are known perfectly, and the only source of bias is the RON. Images of galaxies contain a residual shape measurement error 
\begin{equation}
\Delta \eta^R (\rho_i,\tau_i) = \frac{N_{\rm Tr}}{N^{\rm max}_{\rm Tr}}\sum_i \rho_i f^{\rm res}(\tau_i)\;,
\label{Eq:deltaeta_ref}
\end{equation}
where $\Delta \eta$ may refer to either $\Delta \chi_1$ or $\Delta R^2/R^2$; $N_{\rm Tr}$ is the `serial'/`horizontal' distance (in pixels) of the object to the readout amplifier and $N^{\rm max}_{\rm Tr}$=2099\,pixels is the maximum number of serial transfers given the detector design; the function $f^{\rm res}$ has the  form shown in Eq.~(14) of \cite{Israel15} and coefficients listed in rows 7 and 8 of Table~1 in \cite{Israel15}.
The linear prefactor reflects the increasing number of transfers (hence increasing number of encountered traps) traps for images galaxies farther from the readout amplifier. 
Finally, to model the non-deterministic filling history of traps along the readout direction, which can perturb CTI trailing, we add Gaussian noise to the scaled $\Delta\eta$ in each exposure, with zero mean and 5\% standard deviation. This was not considered in the analysis of \cite{Israel15}, and we find its effect to be negligible.

In the perturbed case, we also propagate uncertain knowledge of the model parameters (e.g.\ accumulated radiation dose) at each point in time. 
Since all our galaxies have the same flux, we introduce model parameter errors in traps' assumed density $\Delta\rho_i$ and release time $\Delta\tau_i$. In addition to errors from back-clocking the RON, as in the reference case, \citet{Israel15} found that model parameter errors introduce a bias
\begin{align}
\Delta \eta_{\rm per} &= \frac{N_{\rm Tr}}{N^{\rm max}_{\rm Tr}}\sum_i \rho_i f^{\rm res}(\tau_i) \\ 
&+ \frac{N_{\rm Tr}}{N^{\rm max}_{\rm Tr}} \sum_i{\left[\rho_i f^{\rm deg}(\tau_i) - (\rho_i+\Delta\rho_i)  f^{\rm deg}(\tau_i+\Delta\tau_i)\right]}\;,\label{Eq:deltaeta_pert}
\end{align}
where the function $f^{\rm deg}$ provides the change (`degradation') in shape parameters because of CTI, without mitigation, as a function of the model parameters. Its functional form is shown in Eq.~(14) of \cite{Israel15} and it uses coefficients listed in rows 3 and 4 of Table~1 in \cite{Israel15}. The difference of this function evaluated at ($\rho+\Delta\rho$, $\tau+\Delta\tau$) from the same function at ($\rho$,$\tau$) in Eq.~(\ref{Eq:deltaeta_pert}) is a reflection of how the iterative mitigation of CTI is, at its root, equivalent to an additional degradation of the images, similar to that caused by CTI, but applied in the opposite sense (hence the name back-clocking). We note that Eq.~(\ref{Eq:deltaeta_pert}) is equal to Eq.~(17) in \cite{Israel15}. Half-way through the mission, both terms account for roughly equal levels of residual.

To assign values to the biases in model parameters, we adopt a constant bias $\Delta\tau_i=1\%$ in the release time parameters, and $\Delta\rho_i$ drawn from a Gaussian distribution with zero mean (average bias is zero in this parameter) and standard deviation of 1\% (over the true value of $\rho$ at each time). They are both conservative, in the sense that they could be derived from {\it Euclid} calibration each day (Nightingale et al.\ in prep.), but $\tau_i$ are likely to be constant for the entire mission and $\rho_i$ smoothly increasing, so errors could be reduced by iterative calibration. They therefore do not necessarily reflect the ultimately achievable  uncertainty in the model parameters, but are useful as reference values.

\begin{figure}
\centering
\includegraphics[width=\columnwidth]{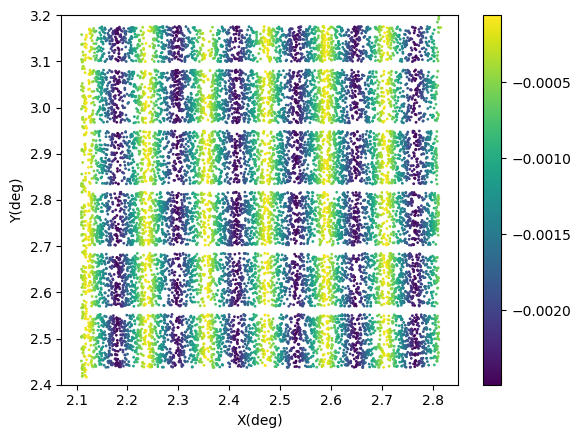}
\caption{A random selection of galaxies are shown representing the pattern of the induced polarisations owing to imperfect CTI mitigation in one field of view. The biases are larger with distance from the readout nodes on either side of the CCDs. Note that we have considered biases only in the serial direction.}
\label{fig:cti_fov}
\end{figure}

Fig.~\ref{fig:cti_fov} shows the pattern of induced biases due to imperfect CTI mitigation for a random selection of galaxies in one FoV.

\end{appendix}

\end{document}